\documentclass{lhep}
\journal{LHEP-431}
\vol{2023}

\usepackage{bbm}
\usepackage{subfigure}
\usepackage{amsmath}
\usepackage{amssymb}
\usepackage{multirow}
\usepackage{dcolumn}
\usepackage{bm}
\usepackage{float}
\usepackage{tabularx}

\begin{document}

\title{Intermediate-Distance String Effects in Wilson Loops\\ via Boundary Action}

\author{M. N. Khalil*, \auno{1,2,3,4} A. M. Khalaf , \auno{4} A. Bakry, \auno{5} M. Deliyergiyev,\auno{6}, A. A. Galal, \auno{4}\\ M. Kotb, \auno{4} M. D. Okasha, \auno{4} and G. S. M. Ahmed, \auno{4}}

\address{$^1$ Department of Mathematics, Bergische Universit\"at Wuppertal, Wuppertal 42097, Germany}
\address{$^2$ Computation-based Science Research Center, Cyprus Institute, Nicosia 2121, Cyprus}
\address{$^3$ Department of Physics, University of Ferrara, Ferrara 44121, Italy}
\address{$^4$ Faculty of Sciences, Department of Physics, Al Azhar University, Cairo 11651, Egypt}
\address{$^5$ Institute of Modern Physics, Chinese Academy of Sciences, Gansu 730000, China}
\address{$^6$ Department of Nuclear and Particle Physics, University of Geneva, CH-211, Switzerland}


\begin{abstract}
The density profile of the QCD flux tube is investigated within the framework of the L\"uscher-Weisz (LW) string action with two boundary terms.  The transverse action profile and potential between static quarks are considered using Wilson's loop overlap formalism at zero temperature in SU(2) Yang-Mills theory. We find the predictions of the LW string matching the lattice data for the width of the energy-density and $Q\bar{Q}$ potential up to a small color-source separation of $R=0.32$\,fm.
\end{abstract}

\maketitle


\section{Introduction}\label{intro}

The confinement of quarks is a fundamental property of quantum chromodynamics (QCD) and strong interactions. Despite extensive research efforts to provide mechanisms of the quark confinement based on the degrees of freedom of QCD, there is no sound analytical construction for the phenomenon of confinement starting from fundamental principles.

The Monte-Carlo calculations of QCD path integrals can unambiguously probe the confinement property on first principles basis. Computer simulations of the confinement potential of the infinitely heavy quark-antiquark pair $Q\bar{Q}$ revealed its linear increasing characteristic~\cite{Creutz:1980zw}.

The linear increase aspect of the potential between two static $Q\bar{Q}$ in the IR region is believed to be a manifestation of a binding gluonic flux tube that assumes a stringlike structure~\cite{Bali:1994de, Haymaker:1994fm, Cea:1995zt,Okiharu:2003vt,Bakry:2017jna,Bakry:2014gea, Bakry:2020xcu}.

String formation is not an uncommon phenomenon among strongly correlated systems. Following the roughening transition~\cite{PhysRevB.78.024510,2007arXiv0709.1042K,Nielsen197345,Lo:2005xt}, the relevant systems show aspects that admit an effective string description. The effective string action is a low-energy effective field theory~\cite{PhysRev.177.2247} which forms a tool for identifying a set of infrared (IR) observables. These predictions can unambiguously be probed in the numerical outcomes of the Monte-Carlo simulations.

In confining gauge models, the effective string theory (EST) predicts~\cite{Luscher:2002qv} a subleading universal correction, proportional to the inverse of string length $\dfrac{1}{R}$, to the linearly rising potential. This is the universal L\"uscher term that has been verified on the lattice simulations of several gauge models~\cite{Juge:2002br, HariDass2008273, Caselle:2016mqu,caselle-2002,Pennanen:1997qm,Brandt:2016xsp}.

Along with the string signatures to the static potential, the EST predicts a logarithmic widening of the energy-density width when the color sources are pulled apart~\cite{Luscher:1980iy}. Lattice simulations~\cite{Juge:2002br,Caselle:1995fh,Bonati:2011nt,HASENBUSCH1994124,Caselle:2006dv,Bringoltz:2008nd,Athenodorou:2008cj,HariDass:2006pq,Giudice:2006hw,Luscher:2004ib,Caselle:2010zs} of several confining gauge groups have confirmed the logarithmic growth property at large enough color source separation.

Despite the successful predictions of the (noninteracting) EST at large color source separations, the analysis of the lattice numerical data for $Q\bar{Q}$ potential and the broadening profile revealed substantial deviations~\cite{PhysRevD.82.094503, Bakry:2010sp}  in the intermediate-distance region at low and high temperatures in addition to the excited spectrum.

Numerous accurate numerical simulations were prompted \cite{Caselle:2004jq, Caselle:2004er} to resolve departures from the free EST by considering the impact that higher-order terms of the LW effective action might have. Other string properties, such as the existence of massive modes owing to the string's stiffness~\citep{Ambjorn:2014rwa,POLYAKOV19,Kleinert:1986bk}, were conjectured~\cite{Ambjorn:2014rwa,POLYAKOV19,Kleinert:1986bk,German:1989vk,Viswanathan:1988ad} and found relevant to compact QED~\citep{Caselle:2016mqu, Caselle:2014eka}, analysis at high temperatures~\cite{bakry2020qcd, Bakry:2020ebo}, and excited energy spectrum~\cite{Brandt:2017yzw}.

Apart from these fine structures of the strings, it has been proposed that the string boundaries which are either two Polyakov lines or Wilson loops can significantly modify the static potential and width profile. This is formally implemented by a surface term in the action of an open string, which appears at derivative orders respecting relativistic invariance.

In fact, the Lorentz-invariant boundary corrections~\cite{Aharony:2010db} to the static $Q\bar{Q}$ potential have demonstrated viability in the analysis of the correlators of both Wilson and Polyakov loop~\cite{Bakry:2020ebo,Brandt:2017yzw,Bakry:2020flt, Brandt:2010bw,Billo:2012da}. The boundary corrections to the string potential recently provided the rationale for the deviations among predictions of the EST and the numerical outcomes~\cite{Billo:2011fd, Billo:2012da}. Moreover, the boundary term modifications to the excited spectrum ~\cite{Brandt:2017yzw, Brandt:2010bw} gave an account for the well-known deviations at relatively large color separation distances.

The consequence of the inclusion of boundary terms in the action is also to correct the characteristic logarithmic broadening of the string's energy width \cite{Bound}. However, a confrontation of the numerical lattice data at zero temperature with these theoretical predictions remains to be addressed.

In this paper, we consider both modifications to the mean-square width \cite{Bound} and the static potential \cite{Billo:2012da} by virtue of two boundary terms, in the order of fourth and sixth derivatives~\cite{Bakry:2020flt,Bound}, in the action. We aim to investigate these analytic predictions with a set of lattice data corresponding to a static meson at different values of lattice coupling.

Section~\ref{sec:LW} lays out the formalism derived from the LW string to estimate the corrections to the $Q\bar{Q}$ and the width of the flux-tubes. In Section~\ref{sec:results}, we confront the numerical behavior of each lattice observable with the LW string prediction. We provide concluding remarks in Section~\ref{sec:concl}.

\section{The Potential and Width of LW\\ String with Boundary Terms}\label{sec:LW}
The conjecture that the Yang-Mills (YM) vacuum admits the formation of a very thin stringlike object~\cite{nambu1974strings} has its origins in the context of the linear rise property of the confining potential between color sources. The formation of a stringlike condensate in the Yang-Mills vacuum spontaneously breaks the translational and rotational symmetries of the QCD vacuum. By virtue of the Goldstone theorem~\cite{Low:2001bw}, the symmetry breaking produces massless transverse modes of Goldstone bosons (GB) of $(d-2)$ dimension.

A string action describing the massless Goldstone modes can be introduced as the derivative expansion of collective string coordinates respecting Poincare and parity invariance. One particular form of this action is the LW action~\cite{Luscher:2002qv} which is gauge-fixed to the physical gauge $X^{1}=\zeta_{0}$ and $X^{4}=\zeta_{1}$, and restricting the string fluctuations to the transverse directions of the worldsheet ${\cal C}$. Invariance under the parity transform would keep only the terms of an even number of derivatives. The L\"uscher and Weisz~\cite{Luscher:2002qv} action up to a six-derivative term reads as
\begin{equation}
{\footnotesize
\begin{split}
 &S^{\rm{LW}}[X]\\
 &=\sigma_{0} A+\dfrac{\sigma_{0}}{2} \int d^2\zeta \left[\left(\dfrac{\partial \mathbf{X}}{\partial \zeta_{\alpha}} \cdot \dfrac{\partial \mathbf{X}}{\partial \zeta_{\alpha}}\right)+ \kappa_2 \left(\dfrac{\partial \mathbf{X}}{\partial \zeta_{\alpha}} \cdot \dfrac{\partial \mathbf{X}}{\partial \zeta_{\alpha}}\right)^2 \right.\\
  &\hphantom{\ \ S^{\rm{LW}}\sigma_{0} A+\dfrac{\sigma_{0}}{2} \int d^2\zeta}
  +\kappa_3 \left(\dfrac{\partial \mathbf{X}}{\partial \zeta_{\alpha}} \cdot \dfrac{\partial \mathbf{X}}{\partial \zeta_{\beta}}\right)^2+\kappa_{4}\left(\dfrac{\partial}{\partial \zeta_{\alpha}}\dfrac{\partial}{\partial \zeta_{\beta}}X^{i}\right) \left(\dfrac{\partial X^{j}}{\partial \zeta_{\gamma}}\dfrac{\partial X^{j}}{\partial \zeta_{\beta}}\right)\\
  &\hphantom{\ \ S^{\rm{LW}}\sigma_{0} A+\dfrac{\sigma_{0}}{2} \int d^2\zeta}
  +\left.\kappa_5 \left(\dfrac{\partial \mathbf{X} }{\partial \zeta_{\alpha}} \cdot \dfrac{\partial \mathbf{X}}{\partial \zeta_{\alpha}}\right)^{3}+\kappa_{6}\left(\dfrac{\partial \mathbf{X} }{\partial \zeta_{\alpha}} \cdot \dfrac{\partial \mathbf{X}}{\partial \zeta_{\alpha}}\right)\left(\dfrac{\partial X^{j}}{\partial \zeta_{\beta}}\dfrac{\partial X^{j}}{\partial \zeta^{\gamma}}\right)\right]\\
  &\quad+\theta_{0} \int d^2\zeta \sqrt{g}~\mathcal{R} +\alpha_{0} \int d^2 \zeta \sqrt{g}~\mathcal{K}^2+S_{b},
\end{split}
\label{LWaction}}
\end{equation}
and the vector $X^{i}(\zeta_{0},\zeta_{1})$ maps the region ${\cal C}\subset \mathbb{R}^{2}$ into $\mathbb{R}^{d}$ where subscript $i=1,2,\dots,d$ denotes the dimension of the embedding background. The map $g^{\alpha,\beta}$ is the two-dimensional induced metric on the worldsheet with local coordinates $\zeta_{\alpha}$ and $\zeta_{\beta}$, indexed with $\alpha,\beta=0,1$. The worldsheet area $A$ and the parameters $\sigma_{0}$, $\alpha_{0} $, and $\theta_{0}$ are the string tension, the rigidity, and the Gauss-curvature parameter, respectively.

The couplings $\kappa_j$; $j=2,\ldots,6$ are effective low-energy parameters. The Lorentz symmetry imposes constraints on the values of these couplings. The evaluations performed in \cite{Luscher:2004ib,Billo:2012da,Aharony:2009gg,Aharony2011,Dubovsky:2012wk} yield coupling values such that $\kappa_{2}=\dfrac{-\sigma}{8}$, $\kappa=\dfrac{\sigma}{4}$, $\kappa_{4}=0$, $\kappa_{5}=\dfrac{\sigma}{16}$, and $\kappa_{6}=\dfrac{-\sigma}{8}$. All the couplings agree with the corresponding coefficients in Nambu-Goto (NG) action expanded up to the six-derivative term~\cite{Dubovsky:2012wk, Billo:2010ix}.

The last two geometrical terms, which encompass the Ricci scalar $\mathcal{R}$ and extrinsic curvature $\mathcal{K}$, are proportional to the intrinsic curvature and the second fundamental form, respectively. In the subsequent discussion, we do not consider the effects of these terms. However, the implications owing to these terms were discussed at a finite temperature~\cite{Bakry:2017fii,axion} and will be considered in a detailed version of this investigation elsewhere.

The action in equation~\eqref{LWaction} encompasses surface/boundary terms $S^{b}$ which arise by virtue of the symmetry breaking at the string's boundaries. The boundary term $S_{b}$ is given by
\begin{equation}
\begin{split}
S_{b}&=\int_{\partial \Sigma} d\zeta_0 \left[
\vphantom{\left( \frac{ \partial \mathbf{X} } {\partial \zeta_1 } \cdot \frac{\partial \mathbf{X} }{ \partial \zeta_1} \right)^{2}}
b_1 \frac{\partial \mathbf{X}}{\partial \zeta_1} \cdot \frac{\partial \mathbf{X}}{\partial \zeta_{1}} + b_2 \frac{\partial \partial \mathbf{X}}{\partial \zeta_{1} \partial \zeta_{0}} \cdot \frac{\partial \partial \mathbf{X}}{\partial \zeta_{1} \zeta_{0}}\right. \\
&\qquad\qquad\quad\ \,+\left.b_3 \left( \frac{ \partial \mathbf{X} } {\partial \zeta_1 } \cdot \frac{\partial \mathbf{X} }{ \partial \zeta_1} \right)^{2}+b_4   \frac{\partial^{2} \partial \mathbf{X}}{\partial \zeta_{0}^2 \partial \zeta_{1}} \cdot \frac{\partial^{2} \partial \mathbf{X}}{\partial \zeta_{0}^2 \partial \zeta_{1}}  \right],
\end{split}
\label{boundaryaction}
\end{equation}
where $b_i$ are the couplings~\cite{Luscher:2004ib} of the boundary terms. Consistency with the open-closed string duality~\cite{Luscher:2004ib} implies a vanishing value of the first boundary coupling $b_{1}=0$. Also, the Lorentz-invariance imposes a vanishing value constraint on both $b_{1}=0$ and $b_3=0$. The other two parameters $(b_2,b_4)$ are free nonuniversal parameters anticipated to characterize a given gauge model.

Within the framework of the light-cone quantization, Arvis~\cite{Arvis:1983fp} obtained the exact ground state potential of the Nambu-Goto string action. The Arvis potential reads
\begin{equation}
\label{Pot_NG_Exact}
V^{\rm{NG}}_{\rm{Exact}}(R)= \sigma_{0} R \sqrt{1-\dfrac{2\pi \gamma}{\sigma_{0} R^{2}}},
\end{equation}
with $\gamma=\frac{(d-2)}{24}$.

Expanding the Arvis potential equation~\eqref{Pot_NG_LO} up to the NNLO term yields
\begin{equation}
  V_{{nn\ell o}}(R)=\sigma_{0} R-\dfrac{\pi \gamma}{R}-\dfrac{\pi^{2} \gamma^{2}}{2 \sigma_{0} R^{3}}-\dfrac{\pi^{3} \gamma^{3}}{2 \sigma_{0}^{2} R^{5}}+\mu, \label{Pot_NG_NNLO}
\end{equation}
with $\mu$ defining a certain UV cutoff~\cite{Luscher:2002qv}. As discussed above, since the expansion of the NG action would agree with LW action equation~\eqref{LWaction} up to six-derivative terms, the string potential equation~\eqref{Pot_NG_NNLO} is equivalent to that obtained from the LW action.

In the limit of an infinite cylinder's length $L_{\mathcal{T}}=\dfrac{1}{\mathcal{T}}$, corresponding to the inverse of the temperature scale $\mathcal{T}$ the two consecutive orders
\begin{align}
  V_{{\ell o}}(R)&=\sigma_{0} R- \dfrac{\pi \gamma}{R}+\mu,\label{Pot_NG_LO}\\
  V_{{n\ell o}}(R)&=\sigma_{0} R-\dfrac{\pi \gamma}{R}-\dfrac{\pi^{2} \gamma^{2}}{2 \sigma_{0} R^{3}}+\mu,\label{Pot_NG_NLO}
\end{align}
coincide with the LO L\"uscher potential~\cite{Luscher:2002qv} and the NLO Dietz-Filk formula \cite{PhysRevD.27.2944}.

The second term in equation~\eqref{Pot_NG_LO} is the famed L\"uscher term of the interquark potential. This term signifies a universal quantum effect independent of the gauge model. However, there is no reason to believe that all orders of power expansion are universal among gauge models~\cite{Caselle:2004jq, Giudice:2009di}. For instance, in \cite{Hellerman:2014cba}, it has been shown that the universality feature extends to the NLO term but not the NNLO six-derivative term with its universality expected only among closed string gauge models~\cite{Aharony:2013ipa}.

The Lorentz symmetry is broken by light-cone quantization at any dimension other than $d=3,26$.  Counterterms have to be introduced \cite{Dubovsky:2012sh,Dubovsky:2014fma} for compatible quantization. The first term appears in the same order as that of the NNLO term $-\dfrac{\pi^3(d-26)}{192\pi R^{5}}$; however, this term vanishes for the ground state.

The Arvis potential is tachyonic~\cite{Arvis:1983fp} at a small string length, and the threshold is defined by the convergence radius of the square root expansion $|\dfrac{2\pi \gamma}{\sigma_{0} R^{2}}|<1$. One should recall that the second geometrical term in the LW action equation~\eqref{LWaction} corresponding to extrinsic curvature circumvents the undesirable features of the NG action such as ghosts, tachyons, or an imaginary static potential between quarks when the distance becomes smaller than the critical distance \cite{Ambjorn:2014rwa,axion,POLYAKOV19,Kleinert:1986bk,German:1989vk,Viswanathan:1988ad}.

The potential~\cite{Billo:2012da} of an open Dirichlet string is given from the partition function of Wilson's loop:
\begin{equation}
V(R) =\dfrac{-1}{T} \log{Z}.
\label{2wb2}
\end{equation}

\begin{figure}[t!]
\begin{center}
\includegraphics[width=.33\textwidth]{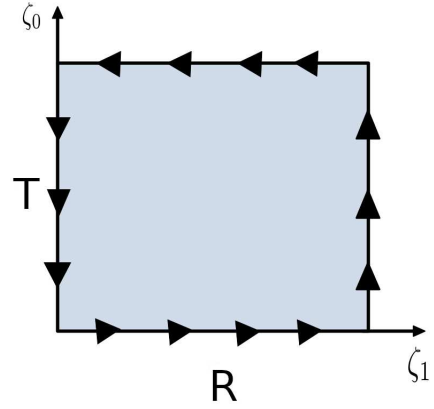}
\caption{\label{WilsonLoop} Rectangular Wilson Loop: the temporal and  spatial length are $T$ and $R$, respectively. The worldsheet coordinates are $\zeta_{0}$ and $\zeta_{1}$. In LGT, the temporal and spatial edges of Wilson loops are the gauge model's links.}
\end{center}
\end{figure}

Figure~\ref{WilsonLoop} shows a conveniently chosen rectangle-shaped Wilson loop circumscribing the spatial-temporal area of $R \times T$. The partition function is the path integral over all string configurations:
\begin{equation}
  Z=\int DX e^{-S^{\rm{LW}}-S_{b_2}-S_{b_4}}.
\label{PF}
\end{equation}

In Wilson loops~\cite{Billo:2011fd,Billo:2012da, Caselle2010pf}, the boundary action $S_{b}$ survives over both the spatial and temporal extents. The boundary action $S_{b_2}$ is then given by
\begin{equation}
  S_{b_{2}}= b_2  \int_{\partial \Sigma_t} d\zeta_0  \frac{\partial \partial \mathbf{X}}{\partial \zeta_{1} \partial \zeta_{0}} \cdot \frac{\partial \partial \mathbf{X}}{\partial \zeta_{1} \zeta_{0}} + b_2 \int_{\partial \Sigma_s} d\zeta_1
\frac{\partial \partial \mathbf{X}}{\partial \zeta_{1} \partial \zeta_{0}} \cdot \frac{\partial \partial \mathbf{X}}{\partial \zeta_{1} \zeta_{0}}.
\label{S2b2Wilson}
\end{equation}

The curves $\partial \Sigma_t$ and $\partial \Sigma_s$ stand for temporal and spatial parts of the Wilson loop, respectively.

Perturbative expansions of the partition function equation~\eqref{PF} around the free action yield
\begin{equation}
 \begin{split}
  Z &=Z_{0} \left(\vphantom{\frac{1}{2}} \left(1-\left\langle S_{b_2} \right\rangle_{\square}-\left\langle S_{b_4} \right\rangle_{\square} \right)\right.\\
  &\hphantom{Z_{0}\ \, \qquad}+\left.\frac{1}{2}  \left\langle \left(S_{b_2}+S_{b_4}+\ldots\right)^2 \right\rangle_{\square} \right)+\ldots,
\end{split}
\end{equation}
where $Z_{0}$ is the partition function of the leading term in LW action $S^{{\rm{LW}}}_{\ell o}$. The direct calculation of the expectation value of the $\langle S_{b} \rangle_{\square}$ entails contributions from both the temporal and spatial parts.

These two contributions give similar formulas of the potential~\cite{Billo:2012da}; however, with the role between the source separation, $R$, and temporal extent, $T$, exchanged,
\begin{equation}
  V^{b_{2}}_{\square}(R,T)=- \dfrac{(d-2)b_2 \pi^{3} }{60 R^{4}}E_{4}\!\left(\frac{iT}{2R}\right)-  \dfrac{(d-2)b_2\pi^{3}  R}{60 T^{5}}E_{4}\!\left(\frac{iR}{2T}\right)\!,
  \label{Pot_Boundb2}
\end{equation}
where $E_{4}$ and  $E_{6}$ are Eisenstein series defined~\cite{abramowitz} according to
\begin{equation}
 \begin{split}
 E_{4}(\tau) = 1+240 \sum_{n=1}^{\infty} \frac{n^3 q^{n}}{1-q^n},\\ E_{6}(\tau) = 1-504\sum_{n=1}^{\infty} \frac{n^5 q^{n}}{1-q^n},
\label{E4}
\end{split}
\end{equation}
respectively, with  $q=e^{i \pi \tau}$ and $\tau=\frac{i T}{2R}$.

Similarly, the boundary potential at the next order $b_4$ would assume the form
\begin{equation}
 V^{b_{4}}_{\square}(R,T)=-\dfrac{ (d-2)b_4\pi^{5}}{126  R^{6}} {E}_{6}\!\left( \frac{iT}{2R}\right)- \dfrac{ (d-2)b_4\pi^{5}R}{126  T^{7}} {E}_{6}\!\left(\dfrac{iR}{2T}\right)\!.
\label{Pot_Boundb4}
\end{equation}
The above expression of the potential due to the next boundary term $V^{b_{4}}$ is derived in~\cite{bakry2020qcd}.

In the following, we put forth the formalism of (MS) width of the string which is another observable that can be probed in the QCD vacuum. The perturbative expectation value of the mean-square (MS) width of the LW string is given by
\begin{equation}
\begin{split}
W_{\rm{LW}}^2(R) &= W_{\ell o}^2(R)- \left\langle \mathbf{X}^{2}S_{n \ell o} \right\rangle\\
&\quad- \left\langle \mathbf{X}^{2} S_{\square,b_2} \right\rangle
- \left\langle \mathbf{X}^{2} S_{\square,b_4}\right\rangle +\ldots,
\label{TwoLoopExpansion}
\end{split}
\end{equation}
where $W_{\ell o}^2$ is the leading-order MS width in $d$ dimension calculated by L\"uscher et al.~\cite{Luscher:1980iy}:
\begin{equation}
  W^{2}_{{\ell o}}\left(R,R_{0}\right) = \frac{d-2}{2\pi\sigma_{0}}\log(R)+R_{0},
\label{W2LO}
\end{equation}
where  $R_{0}$ is the UV cutoff~\cite{Luscher:1980iy,allais,Gliozzi:2010zv}.

The width at the two-loop order of LW action equation~\eqref{LWaction} has been worked out in detail in~\cite{Gliozzi:2010zv, Gliozzi:2010zt}. In the limit of infinite cylinders' height $L_{\mathcal{T}}=\dfrac{1}{\mathcal{T}} \rightarrow \infty$, the NLO term reads
  \begin{equation}
   W^{2}_{n\ell o}(R)= -\frac{\pi W^{2}_{lo}(R)}{4 \sigma_{0} R^2}-\left( \dfrac{(d-2)^2}{96}-\frac{d-2}{16}\right) \dfrac{1}{ \sigma_{0}^{2} R^{2}}.
 \label{SI}
 \end{equation}

Following the same line of reasoning that led to equations~\eqref{Pot_Boundb2} and \eqref{Pot_Boundb4}, the expectation value of the boundary correction to the width of Wilson's loop is evaluated as
\begin{equation}
\begin{split}
  \mathcal{W}^{2}_{\square,b_2}&= \left\langle X^{2} S_{b_2} \right\rangle_{\partial \Sigma_s}+\left\langle X^{2} S_{b_2} \right\rangle_{\partial \Sigma_t}\\
  &=W^{2}_{b_2}(R,T)+W^{2}_{b_2}(T,R),
 \label{2wb2}
\end{split}
\end{equation}
where $W^{2}_{b_2}(R,T)$ is derived in detail in~\cite{bakry2020qcd} and reads as
\begin{equation}
   \begin{split}
   &W^{2}_{b_2}(R,T)\\
   &\quad= \frac{-\pi^{3}  b_2 (d-2)}{16 R^5 \sigma^2_{0}}\left(\frac{1}{3}E_{2}\left(\frac{i T}{R}\right)+2\frac{\vartheta _1^{\prime }}{\vartheta _1}\left( \frac{i\pi T}{4R},q\right)^{\prime} - \frac{91}{6}\right),
   \end{split}
\label{WPb2}
\end{equation}
with $\vartheta_{1}(z,q)$ representing the Jacobi elliptic theta function, where $z \in \mathbb{C}$, and $E_{2}$ denoting the Eisenstein series defined by
\begin{equation}
   E_{2}(\tau) = 1-24 \sum_{n=1}^{\infty} \frac{n q^{n}}{1-q^n}.
\end{equation}
The next-to-leading boundary correction for the Wilson loop of rectangular area $T\times R$~\cite{bakry2020qcd} is, accordingly, given by
\begin{equation}
\begin{split}
  \mathcal{W}^{2}_{\square,b_4}(R,T)=W^{2}_{b_4}(R,T)+W^{2}_{b_4}(T,R),
  \end{split}
\label{2wb4}
\end{equation}
with
\begin{equation}
\begin{split}
  W^{2}_{b_4}(R,T)&=\frac{\pi^5 (d-2)b_4 }{32 R^7 \sigma^2_{0}}
  \left(\vphantom{\left( \frac{i \pi T}{4 R},q\right)^{\prime}}\frac{2}{15}E_{4}\left(\frac{iT}{R} \right)+\frac{1}{3}E_{2}\left(\frac{iT}{R} \right)\right.\\
  &\qquad\qquad\qquad\qquad \left.- \frac{7}{60}-2\frac{\vartheta _1^{\prime }}{\vartheta _1}\left( \frac{i \pi T}{4 R},q\right)^{\prime}-32\right).
\label{WPb4}
\end{split}
\end{equation}
Equations \eqref{WPb2} and~\eqref{WPb4} consistently indicate the same dimension of the couplings $[b_2]=L^3$ and $[b_4]=L^5$ as of equations~\eqref{Pot_Boundb2} and \eqref{Pot_Boundb4}. Scrutinizing the boundary corrections on the lattice entails considering a dimensionless continuum counterpart. The continuum couplings can be conveniently defined \cite{Brandt:2017yzw,Brandt:2010bw, Billo:2012da} through the string tension of a given gauge model as
\begin{equation}
  b_{2}^{c}=b_{2}\sqrt{\sigma_{0}^3};\qquad b_{4}^{c}=b_{2}\sqrt{\sigma_{0}^5}.
\label{Continuum}
\end{equation}
In the following discussion, we ascertain the scaling behavior of the boundary parameters on each lattice coupling.

The leading boundary corrections $b_2$ to the flux tube for either the potential or energy width indicate proportionality to the inverse string length scale at powers $\frac{1}{R^{4}}$ and $\frac{1}{R^{5}}$  equations~\eqref{Pot_Boundb2} and \eqref{2wb2}, respectively. High lattice resolution is, thereof, essential to  detect and disentangle the effects of the two boundary terms $b_{2}$ and $b_{4}$ (of higher inverse power). On top of that, lattice spacing has to be fine in such a way as to avoid systematic errors in the determination of the continuum physics equation~\eqref{Continuum}.

\section{Numerical Results and\\ Discussion}\label{sec:results}
\subsection{Lattice Gauge Theory Data}
Within the following, we compare lattice data calculated by Bali et al.~\cite{Bali:1994de} to the predictions of the string model with boundary action. The lattice gauge theory data represent sets of carefully analyzed correlators corresponding to the $Q\bar{Q}$ potential and energy distribution at different couplings.

The numerical simulation performed by Bali et al.~\cite{Bali:1994de} corresponds to lattices of SU(2) gauge links in pure Yang-Mills theory. The lattices constitute four-dimensional hypercubic Euclidean space-time with periodic boundary conditions. The simulation parameters are summarized in Table~\ref{Tab1} with lattice spacing and string tension in lattice units at each coupling $\beta$. The lattice volumes $L_s^3\times L_t$ are enlisted in Table~\ref{Tab1} with $L_{s}$ and $L_{t}$ defining the spatial and the temporal extents of the 4 torus, respectively.  Temporal links are integrated, and spatial links are smeared (see~\cite{Bali:1994de} for technical details) to improve the signal-to-noise ratio.

\begin{table}[thb!]
\tbl{The string tension and the corresponding lattice spacing at each coupling. The physical scales have been computed from the value \protect{$\sqrt{\sigma}a=440$}\,MeV. \label{Tab1}}{
\begin{tabular}{l|c|c|c}\hline
 & $\beta = 2.50$ & $\beta = 2.635$ & $\beta = 2.74$\\\hline
$L_s^3\times L_t$ &$32^4$&$48^3\times 64$ &  $32^4$ \\
$\sigma_{0}a^{-2}$&0.0350(12)  &0.01458(8)  &.00830(6)\\
$a$ (fm) &0.0826(14)&0.0541(2) &.0408(2)\\\hline
\end{tabular}}
\end{table}

\subsection{The Quark-Antiquark $Q\bar{Q}$ Potential}
The $Q\bar{Q}$ potential can be extracted from the Wilson loop in the limit of large Euclidean time. Wilson loop consists of an ordered product of gauge links with spatial separations $R$ and temporal extent $T$ (see Figure~\ref{WilsonLoop} for the rectangular loops considered here). The $Q\bar{Q}$ pair is then propagated to $\tau=T$. At Euclidean time $\tau=0$, a creation operator
\begin{equation}
\label{create}
\Gamma^{\dagger}_R=Q(0)U(0\rightarrow R)Q^{\dagger}(R),
\end{equation}
where the gauge-invariant spatial link $U(0\rightarrow R)$ is applied to the vacuum state $|0\rangle$ and then finally annihilated by the application of $\Gamma_R$. A spectral decomposition of the Wilson loop reads
\begin{equation}
  \label{expa1}
  \begin{split}
    \langle W(R,T)\rangle&=\frac{1}{\sum_m e^{-E_m T}}\\
    &\quad\times \sum_{m,n}\left|\left\langle m\left|\Gamma_R\right|n,R\right\rangle\right|^2 e^{-V_n(R)T}e^{-E_m (L_{t}-T)}.
\end{split}
  \end{equation}
The ground state potential $V(R)$ can be extracted  in the asymptotic limits $(T\rightarrow\infty)$.
\begin{equation}\label{hh1}
V(R,T)=-\log\left(\frac{\langle W(R,T+1)\rangle}{\langle W(R,T)\rangle}\right).
\end{equation}

The $Q\bar{Q}$ potential data are fitted to theoretical formulas of the LW string potential at the NLO $V_{n\ell o}$ and NNLO $V_{nn\ell o}$ equations~\eqref{Pot_NG_NLO} and~\eqref{Pot_NG_NNLO}, respectively. The number of Wilson loops time slices, at each coupling, is such that the physical length of temporal extents $T \approx 1.7$\,fm.

The cutoff potential is set $\mu$ as a measured fit parameter while keeping the string tension $\sigma_{0} a^{2}$ fixed to the standard values enlisted in Table~\ref{Tab1}. The corresponding returned values of $\chi^{2}$ at each perturbative order are enlisted in Table~\ref{T4_Pot_T09_NG_NLO} for the  three considered lattices of the depicted coupling.

The selected fit intervals of the potential are such that the color source separations are greater than the tachyonic string-length (enlisted for each corresponding $\beta$ in Table~\ref{Tab2}). Even so, we have included fit intervals that are shorter than the tachyonic length by roughly 0.05\,fm and 0.1\,fm. The values of these (listed in Table~\ref{Tab2}) depict a cutoff length defined through $V_{{nn \ell o,n \ell o}}(R_{c})=\mu$ which may imply a threshold of physical potential at the NLO and the NNLO equations~\eqref{Pot_NG_NLO} and \eqref{Pot_NG_NNLO}.

\begin{table}[htb!]
\tbl{The first raw enlists the critical radii at which tachyonic  $Q\bar{Q}$ potential takes place equation~\eqref{Pot_NG_Exact}. The next rows are the radii at the cutoff  perturbative potential $V(R_c)=\mu$ at NLO and NNLO of  LW string  equations~\eqref{Pot_NG_NLO} and \eqref{Pot_NG_NNLO}. \label{Tab2}} {
\begin{tabular}{l|c|c|c}\hline
&   $\beta = 2.50$ & $\beta = 2.635$ & $\beta = 2.74$\\\hline
\small{Order/Radii}&   $R_{c}a^{-1}$  & $R_{c}a^{-1}$  &  $R_{c}a^{-1}$\\\hline
\small{Exact NG/\small{Arvis}} &  3.8678  & 5.9927  &7.9426  \\
\small{NNLO} &  3.3907  & 5.2534  &6.9627 \\
\small{NLO}  &  3.1965  & 4.9526  &6.5641  \\\hline
\end{tabular}}
\end{table}

\begin{table}[htb!]
    \tbl{The $\chi^{2}_{\rm{d.o.f.}}$ values returned from fits to the (NLO) potential $V_{n\ell o}$ equation~\eqref{Pot_NG_NLO} and (NNLO) potential $V_{nn\ell o}$ equation~\eqref{Pot_NG_NNLO}. \label{T4_Pot_T09_NG_NLO}}{%
\begin{tabular}{l|cc}\hline
Fit& $V_{n\ell o}$ & $V_{nn\ell o}$ \\
Interval& $\chi^{2}_{\rm{d.o.f}}$& $\chi^{2}_{\rm{d.o.f}}$ \\\hline
\multicolumn{3}{c}{$\beta=2.5$} \\\hline
$[4,8]$ & 26.6 & 86.0 \\
$[5,8]$ & 8.9 & 16.5 \\
$[6,8]$ & 4.0 & 5.9 \\\hline
\multicolumn{3}{c}{$\beta=2.63$}\\\hline
$[6,12]$ & $10^{3}$ & --- \\
$[7,12]$ & 49.0 & 99.3 \\
$[8,12]$ & 26.8 & 46.6 \\
$[9,12]$ & 9.8 & 15.2 \\
$[10,12]$ & 5.9 & 7.9 \\\hline
\multicolumn{3}{c}{$\beta=2.74$} \\\hline
$[7,16]$ & 55.1 & 187.6 \\
$[8,16]$ & 22.3 & 50.0 \\
$[9,16]$ & 12.2 & 22.9 \\
$[10,16]$ & 5.6 & 9.3 \\
$[11,16]$ & 2.7 & 4.2 \\
$[12,16]$ & 1.7 & 2.5 \\
$[13,16]$ & 0.6 & 0.8 \\\hline
		\end{tabular}
	}
\end{table}

Inspection of Table~\ref{T4_Pot_T09_NG_NLO} reveals that the fit of the pure NG string potential equation~\eqref{Pot_NG_NLO} to the numerical data returns very large values of $\chi^{2}_{\rm d.o.f}$. Despite the reduction in the residuals by the gradual exclusion of short-distance points, the results indicate poor fit all over the considered string length up to color source separation distance $R<0.6$\,fm. The values of $\chi^{2}_{\rm d.o.f}$ returned from the fits of the NNLO $V_{nn\ell o}$  are almost as twice as that of the fits of the NLO $V_{n\ell o}$. However, it seems that both perturbative orders consistently fit well over the interval $R\in[13a,16a]$ of the lattice of fine spacing $a=0.0408(2)$.

The static $Q\bar{Q}$ potential data are fitted to the possibly interesting combination of the leading boundary correction $V^{b_2}$ equation~\eqref{Pot_NG_NLO_Boundb2} together with either of the two perturbative orders  $V_{n\ell o}$ and  $V_{nn\ell o}$ such that
\begin{align}
V_{n\ell o}^{b_2}&=V_{n\ell o}+V^{b_2}_{\square},\label{Pot_NG_NLO_Boundb2}\\
V_{nn\ell o}^{b_2}&=V_{nn\ell o}+V^{b_2}_{\square}.\label{Pot_NG_NNLO_Boundb2}
\end{align}
The returned $\chi^{2}_{\rm{d.o.f.}}$ and $b_2$ from the fits are enlisted in Tables~\ref{T4_Pot_T09_NG_NLO_boundb2beta1}, \ref{T4_Pot_T09_NG_NLO_boundb2beta2}, and \ref{T4_Pot_T09_NG_NLO_boundb2beta3} corresponding to each lattice. The continuum scale parameter $b_{2}^{c}=\sqrt{\sigma_{0}^{3}}b_{2}$ is evaluated at each fit interval.

The boundary-corrected string potential fits show a considerable reduction in $\chi^{2}_{\rm{d.o.f.}}$ across all fit intervals. The fit of the numerical data using the boundary terms $V^{b_2}_{n\ell o}$ produces good $\chi^2_{\rm{d.o.f.}}$ over short fit interval of $R \in [5a,8a]$ at $\beta=2.5$. The potential at NNLO $V^{b_2}_{nn\ell o}$ fits the data nicely on the same fit interval.

Models examination over the lattice of coupling $\beta=2.63$ returns no significant difference between string models $V^{b_2}_{n\ell o}$ and $V^{b_2}_{nn\ell o}$ up to  $R_{\min}=9a$ corresponding to physical distance $R=0.4869$\,fm. However, the fits cease to return good $\chi^{2}_{\rm{d.o.f}}$ for  $V^{b_2}_{nn\ell o}$, at $R < 8a$ and $R < 7a$ for $V^{b_2}_{n\ell o}$. Similarly, at coupling $\beta=2.74$, the stringy behavior set in for source separation $R \in [9a,16a]$, that is, one lattice spacing smaller than fits with NNLO  potential $V^{b_2}_{nn\ell o}$.

\begin{table}[thb!]
\tbl{The $\chi^{2}_{\rm{d.o.f.}}$ values and the boundary parameter $b_2$ together with its continuum counterpart $b_{2}^{c}=\sqrt{\sigma_{0}^{3}}b_{2}$ returned from fits to both NLO $V_{n\ell o}^{b2}$ and NNLO $V_{nn\ell o}^{b2}$ given by equations~\eqref{Pot_NG_NLO_Boundb2} and \eqref{Pot_NG_NNLO_Boundb2}. \label{T4_Pot_T09_NG_NLO_boundb2beta1}}{%
\begin{tabular}{l|ccc}\hline
$\beta=2.5$	&\multicolumn{2}{c}{Fit Parameters}	&\multirow{2}{*}{$\chi^{2}_{\rm{d.o.f.}}$} \\
Fit Interval&$b_{2} $&$b_{2}^{c}~10^{3}$ &\\\hline
 $V_{n\ell o}^{b2}$ &  &          &     \\
$[4,8]$ & 0.25(4) &1.6(3) &18.3\\
$[5,8]$  & 0.16(4)  &1.1(3) & 0.15\\   \hline
$V_{nn\ell o}^{b2}$ &  &          &     \\
$[4,8]$           &0.39(4) & 2.6(3)  &75.0\\
$[5,8]$           &0.22(4) &  1.4(3) &0.47\\\hline
\end{tabular}}
\end{table}

\begin{table}[thb!]
\tbl{Same as Table~\ref{T4_Pot_T09_NG_NLO_boundb2beta1} except the returned $\chi^{2}$ values are from fits to the static $Q\bar{Q}$ potential over a finer-spaced lattice $a=0.01456$\,fm. \label{T4_Pot_T09_NG_NLO_boundb2beta2}}{%
			\begin{tabular}{l|ccc}\hline
$\beta=2.63$&\multicolumn{2}{c}{Fit Parameters}&\multirow{2}{*}{$\chi^{2}_{\rm{d.o.f.}}$}\\
Fit Interval&{$b_{2} $}&{$b_{2}^{c}~10^{3}$}& \\\hline
$V_{n\ell o}^{b2}$ &  &          &     \\
$[6,12]   $       & 0.83(2) & 1.46(4)  & 48.4  \\
$[7,12]  $        & 0.40(3) & 0.70(5)  & 1.3   \\\hline
$V_{nn\ell o}^{b2}$ &  &          &     \\
$[6,12] $    &  1.57(2)  &   2.76(4)       & 248   \\
$[7,12]$     &  0.59(3)  &   1.03(5)       & 6.55      \\
$[8,12]$     &  0.48(5)  &   0.85(9)       & 1.42 \\\hline
\end{tabular}}
\end{table}

\begin{table}[thb!]
\tbl{Same as Table~\ref{T4_Pot_T09_NG_NLO_boundb2beta1}; however, the analyzed data correspond to lattice of spacing $a=0.0408$\,fm. \label{T4_Pot_T09_NG_NLO_boundb2beta3}}{%
			\begin{tabular}{l|ccc}\hline
$\beta=2.74$&\multicolumn{3}{c}{Fit Parameters} \\
Fit Interval&{$b_{2} $}&{$b_{2}^{c}~10^{3}$} &$\chi^{2}_{\rm{d.o.f.}}$\\\hline
$V_{n\ell o}^{b2}$ &           &           &     \\
$[7,16]       $   &0.078(22) &0.059(17) &5.6        \\
$[8,16]      $    &0.9(3)    &0.7(2) & 1.7       \\
$[9,16]     $     &0.8(3)    &0.6(2) & 0.6\\          \hline
$V_{nn\ell o}^{b2}$ &           &           &    \\
$[7,16]    $      &$-$1.5(2) &  $-$1.1(2)  & 74.9    \\
$[8,16]   $       &1.9(3)  &  1.4(2)   & 11.9    \\
$[9,16]  $        &1.5(3)  &  1.1(2)   & 2.7    \\
$[10,16]$         &1.0(3)  & 0.75(2)   & 0.71   \\       \hline
			\end{tabular} }
\end{table}

We explore the prospective effects of incorporating the subsequent boundary correction $V^{b_4}$ equation~\eqref{Pot_Boundb4} to further assess the viability
of higher-order boundary terms. The following two models
\begin{align}
  V_{n \ell o}^{b_2,b_4}&=V_{n\ell o}+V^{b_2}_{\square}+V^{b_4}_{\square},\label{Pot_NG_NLO_Boundb2b4}\\
  V_{nn \ell o}^{b_2,b_4}&=V_{nn\ell o}+V^{b_2}_{\square}+V^{b_4}_{\square},\label{Pot_NG_NNLO_Boundb2b4}
\end{align}
are fitted to the lattice data of the static $Q\bar{Q}$ potential. The returned fit parameters from both models are collected in Tables~\ref{T5_Pot_T09_NG_NLO_boundb2b4beta1}, \ref{T5_Pot_T09_NG_NLO_boundb2b4beta2}, and \ref{T5_Pot_T09_NG_NLO_boundb2b4beta3} for the considered  $\beta$ values. The values of the continuum scale $b_{4}^{c}=\sqrt{\sigma_{0}^{5}}b_4$ are included as well.

We find a subtle difference in the fit behavior between the NLO and  NNLO to diminish within the context of the two-parameter $(b_2,b_4)$ LW string potential. At couplings $\beta=2.5$ and $\beta=2.63$,  the optimal fits are reproduced on the intervals $R \in [4a,8a]$ and $R \in [6a,12a]$ corresponding to a minimal physical length $R=0.3246$\,fm and $R=0.3304$, respectively.

The fits to the boundary-corrected LW string potential are generically reproducing values of the parameter $(b_2,b_4)$ which appear to depend drastically on both the fit interval and lattice coupling $\beta$. As discussed near the end of Section \ref{sec:LW}, we expect the physical relevance of the parameters to scale with the lattice spacing and always preserve the continuum limit. Actually, stability in the values of the continuum parameters $b_{2}^{c}$ is observed for fit over intervals returning residuals around $\chi^{2}_{\rm{d.o.f.}}\simeq 1$ or less. This is evident from the last entries in Table \ref{T4_Pot_T09_NG_NLO_boundb2beta3} corresponding to $\beta=0.263$ and those in Table \ref{T4_Pot_T09_NG_NLO_boundb2beta2}.

 \begin{table}[thb!]
\tbl{The $\chi^{2}_{\rm{d.o.f.}}$ values with the continuum boundary parameters $b^{c}_2=\sqrt{\sigma_{0}^{3}}b_{2}$ and $b_4^{c}=\sqrt{\sigma_{0}^{5}}b_{4}$ and $\mu$ returned from fits to NLO potential $V_{n\ell o}^{b2,b4}$ and NNLO potential $V_{nn\ell o}^{b2,b4}$ given by equations~\eqref{Pot_NG_NLO_Boundb2b4} and  \eqref{Pot_NG_NNLO_Boundb2b4}, respectively.\label{T5_Pot_T09_NG_NLO_boundb2b4beta1}} {
	\resizebox{\columnwidth}{!}{ \begin{tabular}{l|ccccc}\hline
$\beta= 2.5$&\multicolumn{5}{c}{Fit Parameters}\\
Fit Interval&{$b_2 $}&{$b_2^{c}~10^{3}$} &{$b_4$}&{$b_4^{c}~10^{5}$}&{$\chi^{2}_{\rm{d.o.f.}}$} \\\hline
$V_{n\ell o}^{b2,b4}$ &&&&&             \\
$[4,8]$ & 0.003(49) & 0.02(32) & 0.058(5) &1.3(1) &  0.38     \\\hline
$V_{nn\ell o}^{b2,b4}$ &&&&&             \\
$[4,8]$ & 0.003(49) & 0.02(32) & 0.058(5) &1.3(1) &  0.31     \\\hline
\end{tabular}}}
\end{table}

\begin{table}[thb!]
\tbl{Same as Table \ref{T5_Pot_T09_NG_NLO_boundb2b4beta1} except the returned $\chi^{2}$ values are from fits to the static $Q\bar{Q}$ potential over finer-spaced lattice $a=0.01456$\,fm. \label{T5_Pot_T09_NG_NLO_boundb2b4beta2}} {
	\resizebox{\columnwidth}{!}{ \begin{tabular}{l|ccccc}\hline
$\beta= 2.63$&\multicolumn{5}{c}{Fit Parameters}\\
Fit Interval&{$b_2 $} &{$b_2^{c}~10^{3}$} &{$b_4$}&{$b_4^{c}~10^{5}$}&{$\chi^{2}_{\rm{d.o.f.}}$}\\\hline
$V_{n\ell o}^{b2,b4}$ &&&&&   \\
$[6,12]$ & 0.8&1.4(4)   & 0.86(7) & 2.2(2) &0.17 \\\hline
$V_{nn\ell o}^{b2,b4}$ &&&&&   \\
$[6,12]$ & 0.6(2) &1.1(4) & 1.21(7) &3.1(2)  & 0.15     \\\hline
\end{tabular}}}
\end{table}

\begin{table}[thb!]
\tbl{Same as Table~\ref{T5_Pot_T09_NG_NLO_boundb2b4beta1}; however, the analyzed data correspond to lattice of spacing $a=0.0408$\,fm.\label{T5_Pot_T09_NG_NLO_boundb2b4beta3}} {
	\resizebox{\columnwidth}{!}{ \begin{tabular}{l|ccccc}\hline
$\beta= 2.74$&\multicolumn{5}{c}{Fit Parameters}\\
Fit Interval&{$b_2 $}  &{$b_2^{c}~10^{3}$} &{$b_4$}&{$b_4^{c}~10^{5}$}&{$\chi^{2}_{\rm{d.o.f.}}$} \\\hline
$V_{n\ell o}^{b2,b4}$            &           &  &&&   \\
$[7,16]$   & 0.8(3)    &0.6(2) &1.0(6)&0.91(5)   &1.06   \\\hline
$V_{nn\ell o}^{b2,b4}$            &           &  &&&   \\
$[7,16]$ &0.2(3)  &0.1(3) &2.5(1) &1.6(6)  & 0.83   \\\hline
\end{tabular}}}
\end{table}

To further oppose the outcomes in Tables~\ref{T5_Pot_T09_NG_NLO_boundb2b4beta2} and \ref{T5_Pot_T09_NG_NLO_boundb2b4beta3}, we should beware of the small values of $\chi^{2}_{\rm{d.of}} \leq 1$. For instance, when fitting $V_{n\ell o}^{b_2,b_4}$ with $b_{4}=0.76$ and $b_{4}^{c}=1.3$ over the range $R \in [6a,12a]$, the resulting $\chi^{2}=5.9$ is consistent with the uncertainties of the values of $R_{0}$ and $b_{2}$ listed in Table~\ref{T5_Pot_T09_NG_NLO_boundb2b4beta2}. A similar argument applies to the second entry in Table~\ref{T5_Pot_T09_NG_NLO_boundb2b4beta3}, adopting $b_{4}=1.13$; $b_{4}^{c}=1.94$, such that it approaches the continuum parameter $b_{4}^{c}=1.6(6)$ in Table~\ref{T5_Pot_T09_NG_NLO_boundb2b4beta3}, and reproduces good $\chi^{2}=4.8$ where $R_{0}$ and $b_{2}$ are within its uncertainties. Similar considerations hold for the outcomes in Table~\ref{T5_Pot_T09_NG_NLO_boundb2b4beta1}.

\begin{figure*}[htb!]
  \begin{center}
    	\subfigure[]{\includegraphics[scale=0.55]{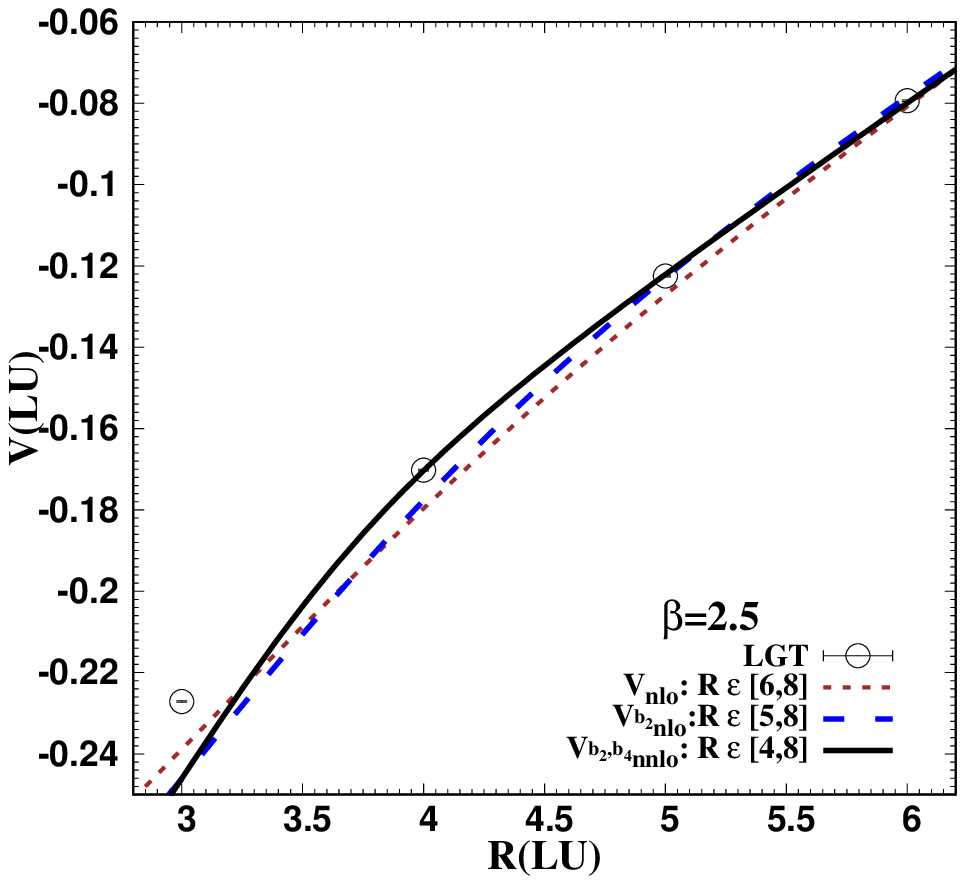}}
   	\subfigure[]{\includegraphics[scale=0.55]{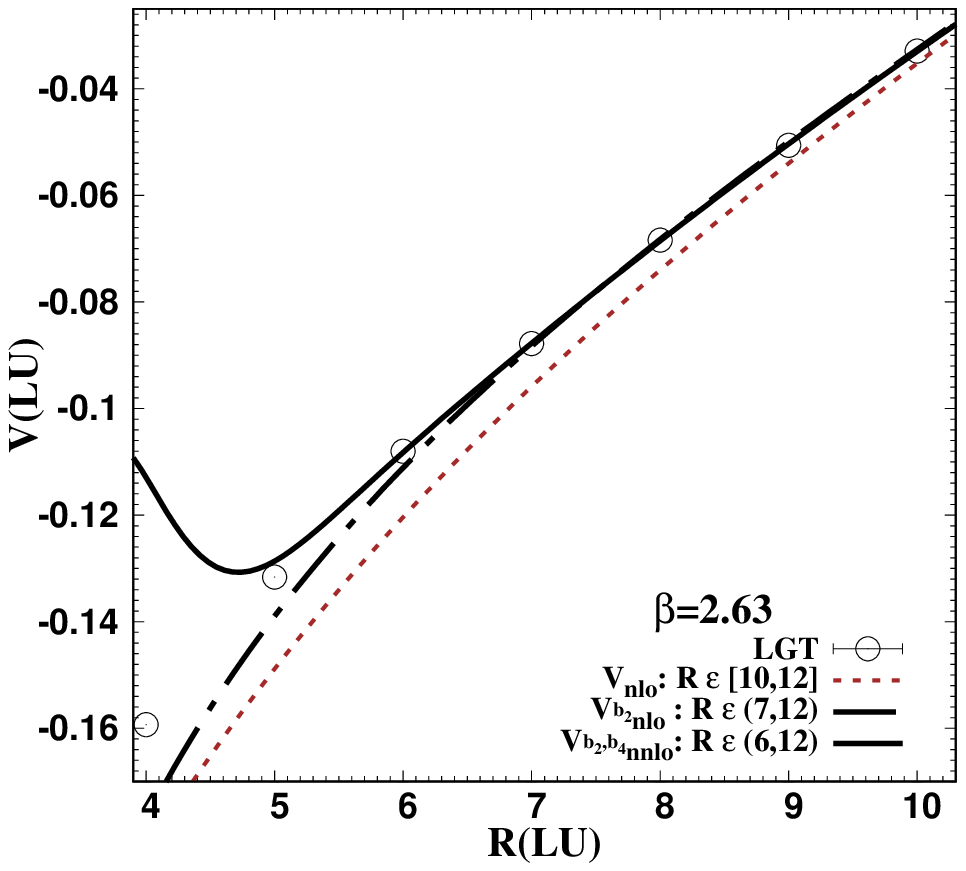}}
        \subfigure[]{\includegraphics[scale=0.55]{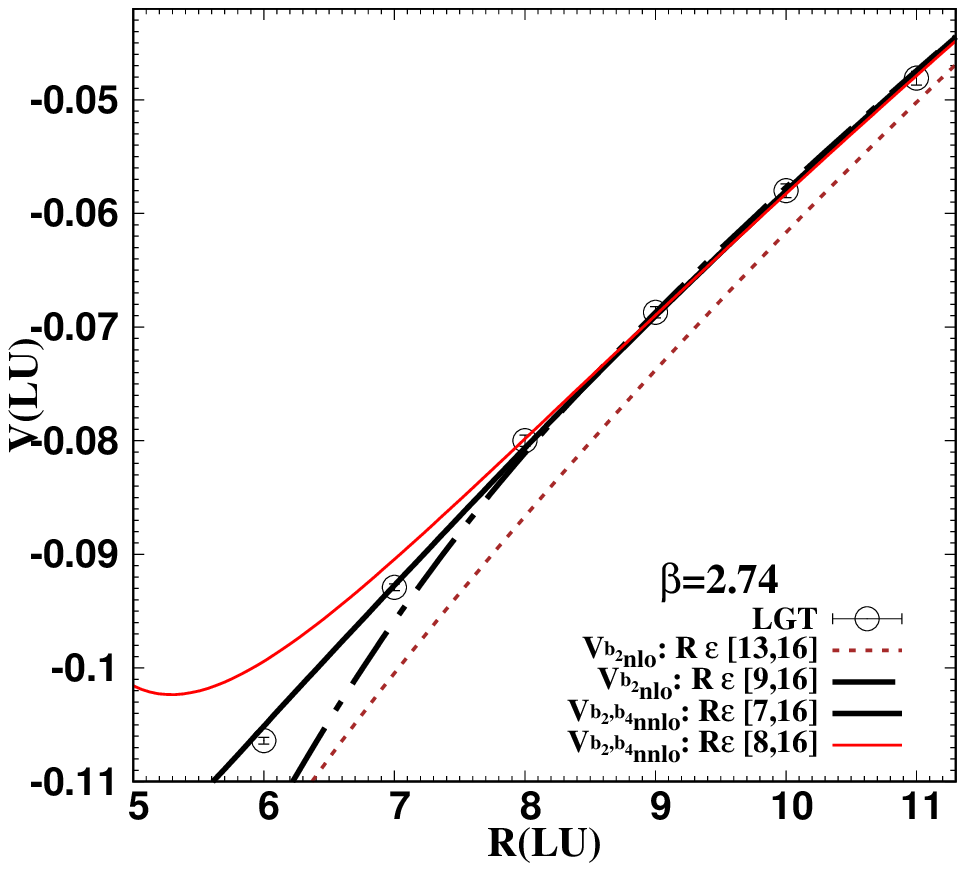}}
\caption{\label{VNG_LoNloB2B4}The quark-antiquark $Q \bar{Q}$ potential data at three different lattice spacing $a$. The dotted and dashed lines correspond to fits of the LW string potential at NLO $V_{n\ell o}$  equation~\eqref{Pot_NG_NLO_Boundb2} and NLO potential  with boundary terms $V^{b2}_{n\ell o}$  equation~\eqref{Pot_NG_NLO_Boundb2}, respectively. The solid line corresponds to best fits to LW string at NNLO with two boundary terms $V^{b2,b4}_{nn\ell o}$  equation~\eqref{Pot_NG_NNLO_Boundb2b4}.}
\end{center}
\end{figure*}

Three plots at each assumed coupling $\beta$ are collected in the panel of Figure~\ref{VNG_LoNloB2B4}. The figures show the lattice data of the potential's $Q\bar{Q}$ together with the lines that best fit each string model prediction. The legend depicts the interval of the best fits.

One clearly observes the improvement in the fit with respect to the LW string with one boundary term $V_{n \ell o}$ compared LW model at NLO $V_{n \ell o}$. The fit of the potential with one boundary term $V_{n \ell o}^{b_2}$ indicates at least three lattice spacing values of the overall improved match. The two-term boundary potential $V_{nn \ell o}^{b_2}$; however, finely corrects at least one lattice spacing value at each lattice coupling.

\subsection{Energy-Density Profile}

To characterize the Euclidean action density on the lattice we utilize a plaquette operator defined by
\begin{equation}
    \mathcal{P}_{\mu\nu}(\boldsymbol\rho)= \left[U_{\mu}(\boldsymbol\rho)U_{\nu}(\boldsymbol\rho+\boldsymbol\mu)U_{\mu}^{\dagger}(\boldsymbol\rho+\boldsymbol\nu)U_{\nu}^{\dagger}(\boldsymbol\rho)\right],
\end{equation}
with the indices $\mu$ and $\nu$ corresponding to Lorentz indices.

The Euclidean action density at position $\boldsymbol\rho$ is given by
\begin{equation}
S(\boldsymbol\rho) =\beta \sum_{\mu<\nu} \left( 1-\frac{1}{3} \mathrm{Re}\,\mathrm{Tr}\,\mathcal{P}_{\mu\nu}(\boldsymbol\rho) \right),
\end{equation}
where $\beta$ is the coupling of Yang-Mills theory. The plaquette $\mathcal{P}_{\mu\nu}$ can be expanded in a power series in the symmetric field strength tensor $F_{\mu\nu}$ such that
\begin{equation}
S(\boldsymbol\rho) = a^{4} \sum_{\mu<\nu} \mathrm{Tr}\, F^{2}_{\mu\nu}(\boldsymbol\rho) +\mathcal{O}\left(a^{2}\right)+ \mathcal{O}\left(a^{2}g^{2}\right)
\end{equation}
with $g^{2}=\dfrac{6}{\beta}$.

%

A dimensionless scalar field characterizing the Euclidean action density distribution in the Polyakov vacuum, i.e., in the presence of color sources can be defined as
\begin{equation}
	\mathcal{C}\left(\boldsymbol{\rho};\mathbf{r}_{1},\mathbf{r}_{2}\right)= \frac{\left\langle \mathcal{P}_{2Q}(\mathbf{r}_{1},\mathbf{r}_{2})\right\rangle
\langle S(\boldsymbol{\rho}) \rangle-\left\langle\mathcal{P}_{2Q}\left(\mathbf{r}_{1},\mathbf{r}_{2}\right) S(\boldsymbol{\rho})\right\rangle }
{\left\langle \mathcal{P}_{2Q}\left(\mathbf{r}_{1},\mathbf{r}_{2}\right)\right\rangle\langle S(\boldsymbol{\rho})\rangle},
	\label{Actiondensity}
\end{equation}
with the vector $\boldsymbol{\rho}$ referring to the spatial position of the energy probe with respect to some origin, and the bracket $\langle\cdots\rangle$ stands for averaging over gauge configurations and lattice symmetries. Other dimensionful definitions of the correlator \eqref{Actiondensity} yield an equivalent representation of the width (see, for example, \cite{Okiharu:2003vt}).


The width of the transverse profile of the action density equation~\eqref{Actiondensity} is extracted from the fits to Gaussian form~\cite{Bali:1994de}, which within the accuracy of the measurements return good $\chi^{2}$ for color source separation  $R>0.2$\,fm.

We discuss the growth in the width of the action density with the increase of the $Q\bar{Q}$ separation according to either of the following three models.

The pure LW string models at the two-loop order:
\begin{equation}
W^2_{1}=W^2_{{(\ell o)}}+W^2_{{(n\ell o)}},
\label{fitmod1}
\end{equation}
in accordance with equations~\eqref{W2LO} and \eqref{SI}. In addition to the MS width of LW string with one and two boundary corrections:
\begin{align}
W^2_{2}&=W^2_{{(\ell o)}}+W^2_{{(n\ell o)}}+W^2_{b_2},
\label{fitmod2}\\
W^2_{3}&=W^2_{{(\ell o)}}+W^2_{{(n\ell o)}}+W^2_{b_2}+W^2_{b_4},
\label{fitmod3}
\end{align}
respectively, in accordance with equations~\eqref{W2LO}, \eqref{SI}, \eqref{2wb2}, and \eqref{2wb4}.

The analysis of the fit behavior of the MS width data is discussed keeping the string tension fixed to the standard values enlisted in Table~\ref{Tab1}. The fits are obtained by solving an extremum problem in the parameter space ${R_0, b_2, b_4}$ such that the least square residuals
 \begin{equation}
\chi^2\left(R_0,b_2,b_4\right)= \sum_{i}\left(\frac{ W^2\left(R_i\right)-W^2_{\text{model}}\left(R_i;R_0,b_2,b_4\right)}{2W\left(R_i\right)e\left(R_i\right)}\right)^2
\label{chisquaredb2b4}
\end{equation}
are minimized. It should be noted that, in equation~\eqref{chisquaredb2b4}, $e(R_i)$ corresponds to the error in the width $W(R_i)$. We consider the lattice data of the MS width of the QCD flux tube at the perpendicular plane in the middle $R/2$ between the color sources~\cite{Bali:1994de}.

The resultant minimum $\chi^{2}_{\rm{d.o.f.}}$ and the corresponding fit parameters $R_{0}$, $b_2$, and $b_4$ from the fits to the lattice data are collected in Tables~\ref{beta25_1}--\ref{beta263_4}.

\begin{table}[thb!]
 \tbl{The returned $\chi^{2}_{\rm{d.o.f.}} $ from the fit of the LW string at NLO  equation~\eqref{fitmod1} to the data of MS width of flux tube, lattice spacing $a=0.0804$\,fm, over two fit intervals $[R_{\min},R_{\max}]$. $R_{0}$ is the UV cutoff parameter equation~\eqref{W2LO}.\label{beta25_1}}{				%
\begin{tabular}{l|c|c}\hline
$\beta = 2.5$ & Fit Range $[4,14]$ & $\chi^{2}_{\rm{d.o.f.}} =48.2$   \\\hline
 Fit~Parameters & Estimate	    & Standard Error \\\hline
 $R_0$   & 3.634     & 0.675146  \\\hline
\end{tabular}}
\end{table}

\begin{table}[thb!]
 \tbl{Same as Table~\ref{beta25_1}; however, the fits are performed on data on the lattice of finer spacing $a=0.0541$\,fm.\label{beta263_1}}{
   \begin{tabular}{l|c|c}\hline
$\beta = 2.63$  & Fit Range $[6,20]$ & $\chi^{2}_{\rm{d.o.f.}} =72.64$  \\\hline
Fit~Parameters             & Estimate	           & Standard Error \\\hline
$R_{0} $     &  12.68             &0.327134  \\\hline
  & Fit Range $[8,20]$ & $\chi^{2}_{\rm{d.o.f.}} =8.25$  \\\hline
$R_{0} $     &   $-$4.74331           &1.04629 \\\hline
 \end{tabular}}
\end{table}

\begin{table}[thb!]
 \tbl{The retrieved $\chi^{2}_{\rm{d.o.f.}}$ of the fits of the MS width of the LW string with one boundary parameter $b_{2}$ equation~\eqref{fitmod2} to the MS width data.\label{beta25_2}}{				%
\begin{tabular}{l|c|c}\hline
$\beta = 2.5$ & Fit Range $[4,12]$ & $\chi^{2}_{\rm{d.o.f.}} = 0.16$   \\\hline
    Fit~Parameters	       & Estimate	    & Standard Error \\\hline
$R_{0}$ & 10.8904 & 15.1857    \\
  $b_2 $ & 0.22468 & 0.07494  \\
    $b_2^{c}~10^{3}$   &1.47121  &0.49069  \\
    $b_4 $            & 0.09324  & 0.03392   \\
     $b_4^{c} ~10^{5}$ & 2.13692 &  0.77725   \\ \hline
\end{tabular}}
\end{table}

\begin{table}[thb!]
 \tbl{Same as Table~\ref{beta25_2}; however, $a=0.0541$\,fm and fit interval commences at $R=3.24$\,fm. \label{beta263_2}}{				%
\begin{tabular}{l|c|c}\hline
$\beta = 2.63$ & Fit Range $[6,20]$       & $\chi^{2}_{\rm{d.o.f.}} = 3.79$   \\\hline
Fit~Parameters	       & Estimate	        & Standard Error \\\hline
  $R_0 $ &	 $-$267.702 & 15.1058  \\
  $b_2 $ &	                        0.86463 & 0.046570     \\	
  $b_2^{c}~10^{3}$ & 1.52210 & 0.081982     \\	\hline
\end{tabular}}
\end{table}

\begin{table}[thb!]
 \tbl{Same as Table~\ref{beta263_2} except that the fit interval commences at $R=4.328$\,fm. \label{beta263_3}}{				%
\begin{tabular}{l|c|c}\hline
$\beta = 2.63$ & Fit Range $[8,20]$       & $\chi^{2}_{\rm{d.o.f.}} =1.09 $   \\\hline
Fit~Parameters	       & Estimate	        & Standard Error \\\hline
  $R_0 $ &$-$268.448 & 43.3449   \\
  $b_2 $                            &0.86710 & 0.14248     \\	
    $b_2^{c}~10^{3}$ &1.52645 & 0.25082     \\	\hline
\end{tabular}}
\end{table}

\begin{table}[thb!]
 \tbl{The retrieved $\chi^{2}_{\rm{d.o.f.}}$ and fit parameters, $b^{c}_{2}$ and $b^{c}_{4}$ of the fits of the MS width of the LW string, with two boundary parameters equation~\eqref{fitmod3}, to the MS width data. The fit range commences at $R=3.24$\,fm. \label{beta263_4}}{				%
\begin{tabular}{l|c|c}\hline
$\beta = 2.63$ & Fit Range $[6,20]$       & $\chi^{2}_{\rm{d.o.f.}} =1.06$   \\\hline
Fit~Parameters	       & Estimate	        & Standard Error \\\hline
$R_0 $ &$-$258.75 & 197.143  \\
  $b_2 $                            &0.8441 & 0.4532 \\
    $b_2^{c}~ 10^{3}$&1.4859 & 0.7976 \\
  $b_4 $            &0.01712 & 0.37605 \\
  $b_4^{c} ~ 10^{5}$ &0.54163 & 1.18920 \\\hline
\end{tabular}}
\end{table}

\begin{figure}[htb!]
   \begin{center}
     \subfigure[]{\includegraphics[scale=0.70]{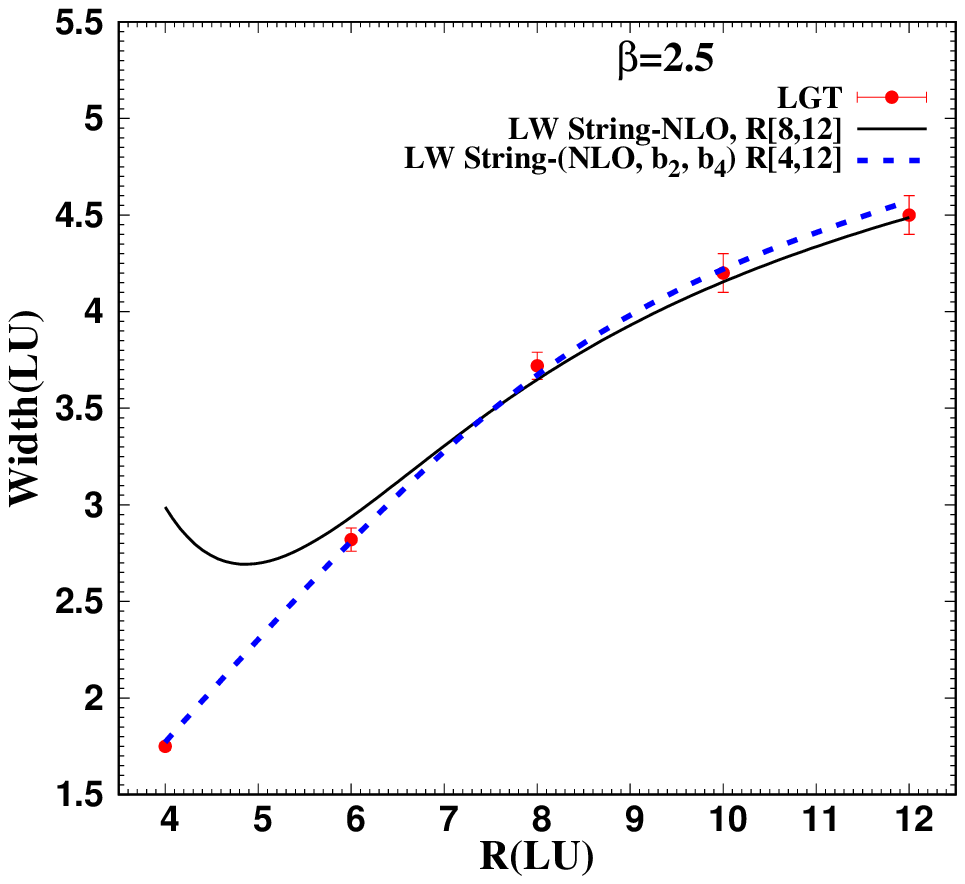}}
     \subfigure[]{\includegraphics[scale=0.70]{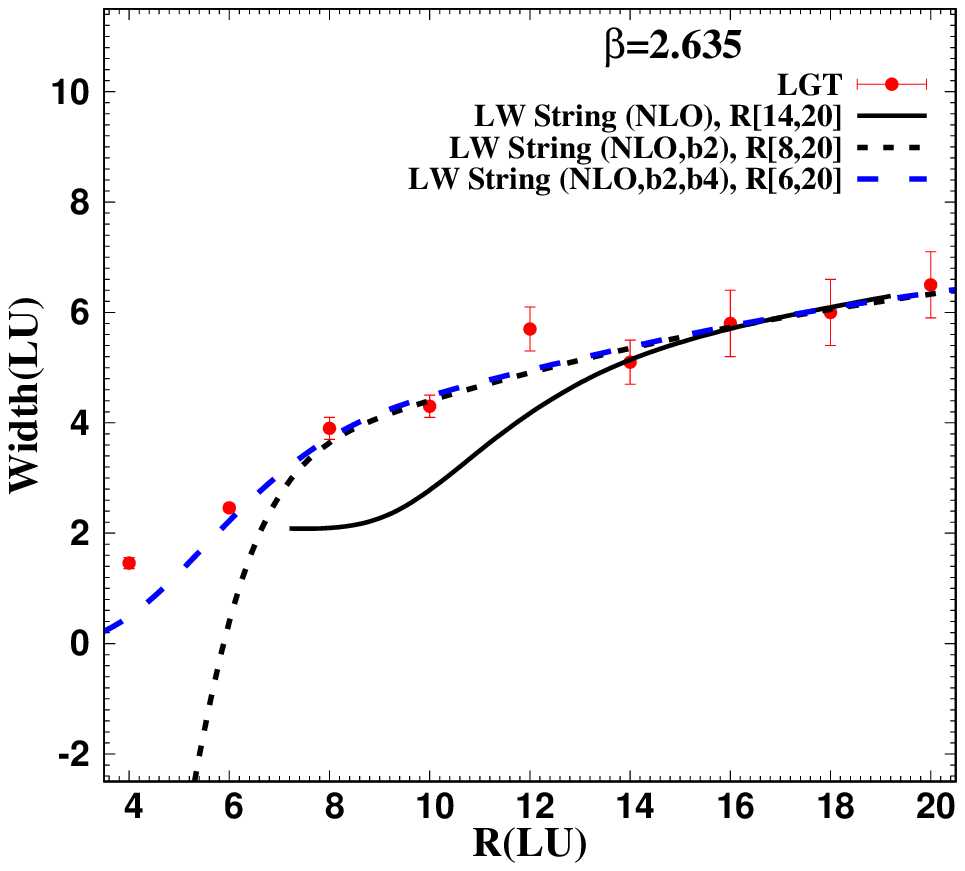}}
\caption{\label{WFit}(a) The data points correspond to the MS width of the flux tube at the middle plane $\dfrac{R}{2}$ at $\beta=2.5$ for the depicted fit ranges. The solid and dotted lines are the fit of the MS width of the LW string at NLO equation~\eqref{fitmod1} and that with boundary term $b_2$  equation~\eqref{fitmod2}. (b) Same as plot (a), except that the MS width is measured on a finer lattice of spacing $a=0.0541$\,fm. The dashed line corresponds to fits to the LW string with two boundary terms $b_{2}$ and $b_{4}$ equation~\eqref{fitmod3}.}
\end{center}
 \end{figure}

The large $\chi^{2}_{\rm{d.o.f.}}$ values in Tables \ref{beta25_1} and \ref{beta263_1} reflect the poor fits received from the NLO approximation equation~\eqref{fitmod1} of the pure LW string model (without boundary corrections). The plots in Figure~\ref{WFit}(b) show that this string model must have considerable source separations $R \geq 0.6$\,fm before it can best match the numerical data. These findings imply that the pure LW string action fails to adequately integrate the subtle characteristics of QCD flux tubes at close ranges.

The parameters retrieved by fitting the width data to the LW string with two boundary terms corresponding to equation~\eqref{fitmod3} are collected in Table \ref{beta25_2}. The results demonstrate the significant reduction of residuals $\chi^{2}_{\rm{d.o.f}}$ compared to that retrieved from the fits of NLO model equation~\eqref{fitmod1}. The fits are  even better than that of the one boundary parameter equation~\eqref{fitmod2} returning $\chi^{2}_{\rm{d.o.f}}=2.5$. The plots in Figure~\ref{WFit}(a) show how well the lattice data to source separations $R \geq 0.2478$ and the LW string with $b_2$ and $b_4$  match each other.

The fits of the LW string model with $b_2$ equation~\eqref{fitmod2} over the finer lattice, $a=0.054$\,fm and $\beta=2.63$, provide optimal flux tube width over color separation interval $R \in [8a,20a]$, Tables~\ref{beta263_2}--\ref{beta263_4}. The corresponding plot in Figure~\ref{WFit}(b) demonstrates a good match with the lattice data up to a minimal string length of $R=0.4328$\,fm.

Considering fit intervals with smaller color source separations $R\in[6a,20a]$ at coupling $\beta=0.263$, which is still above the tachyonic threshold radius $R_c$, the corresponding returned results in Table~\ref{beta263_4} reflect a good $\chi^{2}_{\rm{d.o.f}}=1.06$ from the fit of the two-parameter LW string $(b_2,b_4)$. Figure~\ref{WFit}(b) displays a better match of one lattice spacing, corresponding to the minimal length $R=0.3246$\,fm than fits using the $b_2$ term.

The characteristics of the energy profile of the QCD string should be understood in the context of the complementary IR observable, namely, the ground state potential $Q\bar{Q}$. Despite the comparatively higher uncertainty in the action density equation~\eqref{Actiondensity} than in the  $Q\bar{Q}$ potential, the observation of the string over intervals commencing from the same source separation demonstrates the relevance of the boundary action to the physics of the confining flux tube.

More regularities among the fit parameters of the boundary action are observed when comparing the fits of the $Q\bar{Q}$ potential or the energy width. For example, the fits of the energy density on the interval $R\in[8a,20a]$ shown in Table~\ref{beta263_3} produce $b_2=0.87(0.14)$ which is the same produced from the fits of the $Q\bar{Q}$ potential in Table~\ref{T4_Pot_T09_NG_NLO_boundb2beta3} $R\in[8a,16a]$ with $b_2=0.8(3)$. At the coupling $\beta=2.5$, opposing the parameters in Tables \ref{T4_Pot_T09_NG_NLO_boundb2beta1} and \ref{T5_Pot_T09_NG_NLO_boundb2b4beta1} with those in Table \ref{beta25_2} shows comparable continuum numerical values within the statistical uncertainties.

\section{Conclusion}\label{sec:concl}
In this paper, the quark-antiquark ($Q\bar{Q}$) potential and energy profile of a static meson are compared to the theoretical predictions based on the L\"uscher-Weisz (LW) string with two boundary terms. Link-integrated Wilson loop correlators with optimal overlap with the ground state~\cite{Bali:1994de} are discussed.

At intermediate distances, we detect signatures of the two boundary terms of the L\"uscher-Weisz (LW) string~\cite{Billo:2012da,bakry2020qcd} in the Monte-Carlo data of the static $Q\bar{Q}$ potential. The boundary-corrected string model extends the region of validity of the string for color source separation $R \geq 0.37$\,fm using one boundary term $b_2$, and $R \geq 0.32$\,fm for two boundary terms $(b_2,b_4)$.

The boundary-corrected width \cite{Bound} with one boundary term $b_2$ for the (LW) action reduces the residuals of the fits to the lattice data of the action density, with good fits obtained for string length $R \geq 0.5$\,fm. The inclusion of the second Lorentz-invariant boundary term $b_4$ discloses a good match with the LGT data for color source separation $R \geq 0.32$\,fm


\section*{Acknowledgments}
The authors would like to thank G. Bali for his useful and encouraging comments. We also acknowledge  suggestions and careful revision of the manuscript provided by the LHEP reviewers.

\bibliographystyle{unsrt}

\begin{thebibliography}{}

\end{thebibliography}


\begin{thebibliography}{10}

\bibitem{Creutz:1980zw}
M.~Creutz.
\newblock \textit{Monte carlo study of quantized SU(2) gauge theory}.
\newblock {\em Phys. Rev.}, \textbf{D21}:2308--2315, 1980.

\bibitem{Creutz:1983ev}
Michael Creutz, Laurence Jacobs, and Claudio Rebbi.
\newblock \textit{Monte carlo computations in lattice gauge theories}.
\newblock {\em Phys. Rept.}, 95:201, 1983.

\bibitem{Creutz:1980hb}
Michael Creutz.
\newblock \textit{Monte carlo study of renormalization in lattice gauge
  theory}.
\newblock {\em Phys. Rev.}, \textbf{D23}:1815, 1981.

\bibitem{Creutz:1980vq}
Michael Creutz.
\newblock \textit{Quark confinement}.
\newblock {\em Madison H.E. Physics}, \textbf{296}, 1980.

\bibitem{Fukugita:1983du}
M~Fukugita and T~Niuya.
\newblock The distribution of chromoelectric flux in su (2) lattice gauge
  theory.
\newblock {\em Physics Letters B}, 132(4-6):374--378, 1983.

\bibitem{Flower:1985gs}
Jon~W Flower and Steve~W Otto.
\newblock The field distribution in su (3) lattice gauge theory.
\newblock {\em Physics Letters B}, 160(1-3):128--132, 1985.

\bibitem{Wosiek:1987kx}
J~Wosiek and Richard~W Haymaker.
\newblock Space structure of confining strings.
\newblock {\em Physical Review D}, 36(10):3297, 1987.

\bibitem{Sommer:1987uz}
Rainer Sommer.
\newblock Chromoflux distribution in lattice qcd.
\newblock {\em Nuclear Physics B}, 291:673--691, 1987.

\bibitem{DiGiacomo:1989yp}
Adriano Di~Giacomo, Michele Maggiore, and {\v{S}}tefan Olejn{\'\i}k.
\newblock Evidence for flux tubes from cooled qcd configurations.
\newblock {\em Physics Letters B}, 236(2):199--202, 1990.

\bibitem{DiGiacomo:1990hc}
Adriano Di~Giacomo, Michele Maggiore, and {\v{S}}tefan Olejn{\'\i}k.
\newblock Confinement and chromoelectric flux tubes in lattice qcd.
\newblock {\em Nuclear Physics B}, 347(1-2):441--460, 1990.

\bibitem{Bali:1994de}
G.~S. Bali, K.~Schilling, and C.~Schlichter.
\newblock {Observing long color flux tubes in SU(2) lattice gauge theory}.
\newblock {\em Phys. Rev.}, D51:5165--5198, 1995.

\bibitem{Haymaker:1994fm}
Richard~W Haymaker, Vandana Singh, Yingcai Peng, and Jacek Wosiek.
\newblock Distribution of the color fields around static quarks: Flux tube
  profiles.
\newblock {\em Physical Review D}, 53(1):389, 1996.

\bibitem{Cea:1995zt}
Paolo Cea and Leonardo Cosmai.
\newblock Dual superconductivity in the su (2) pure gauge vacuum: A lattice
  study.
\newblock {\em Physical Review D}, 52(9):5152, 1995.

\bibitem{Okiharu:2003vt}
F~Okiharu and RM~Woloshyn.
\newblock A study of colour field distributions in the baryon.
\newblock {\em Nuclear Physics B-Proceedings Supplements}, 129:745--747, 2004.

\bibitem{Bakry:2017jna}
A.~Bakry, X.~Chen, M.~Deliyergiyev, A.~Galal, S.~Xu, and P.~M. Zhang.
\newblock {Mesonic String of Diquark-Quark Configuration at Finite
  Temperature}.
\newblock {\em PoS}, Hadron2017:211, 2018.

\bibitem{Bakry:2014gea}
Ahmed~S. Bakry, Xurong Chen, and Peng-Ming Zhang.
\newblock {Y-stringlike behavior of a static baryon at finite temperature}.
\newblock {\em Phys. Rev.}, D91:114506, 2015.

\bibitem{Bakry:2020xcu}
A.~S. Bakry, M.~A. Deliyergiyev, A.~A. Galal, and A.~M. Khalaf.
\newblock {Strings of diquark-quark (QQ)Q baryon before phase transition}.
\newblock 2 2020.

\bibitem{PhysRevB.78.024510}
Mark~G. Alford and Gerald Good.
\newblock Flux tubes and the type-i/type-ii transition in a superconductor
  coupled to a superfluid.
\newblock {\em Phys. Rev. B}, 78(2):024510, Jul 2008.

\bibitem{2007arXiv0709.1042K}
K.~{Kasamatsu} and M.~{Tsubota}.
\newblock {Quantized vortices in atomic Bose-Einstein condensates}.
\newblock {\em ArXiv e-prints}, September 2007.

\bibitem{Nielsen197345}
H.B. Nielsen and P.~Olesen.
\newblock Vortex-line models for dual strings.
\newblock {\em Nuclear Physics B}, 61:45 -- 61, 1973.

\bibitem{Lo:2005xt}
Amy~S. Lo and Edward~L. Wright.
\newblock {Signatures of cosmic strings in the cosmic microwave background}.
\newblock 2005.

\bibitem{PhysRev.177.2247}
Curtis~G Callan~Jr, Sidney Coleman, Julius Wess, and Bruno Zumino.
\newblock Structure of phenomenological lagrangians. ii.
\newblock {\em Physical Review}, 177(5):2247, 1969.

\bibitem{Battelli:2019lkz}
Nico Battelli and Claudio Bonati.
\newblock Color flux tubes in s u (3) yang-mills theory: An investigation with
  the connected correlator.
\newblock {\em Physical Review D}, 99(11):114501, 2019.

\bibitem{Caselle:1995fh}
M.~Caselle, F.~Gliozzi, Ulrika Magnea, and S.~Vinti.
\newblock {Width of long colour flux tubes in lattice gauge systems}.
\newblock {\em Nucl. Phys.}, B460:397--412, 1996.

\bibitem{Bonati:2011nt}
Claudio Bonati.
\newblock {Finite temperature effective string corrections in (3+1)D SU(2)
  lattice gauge theory}.
\newblock {\em Phys.Lett.}, B703:376--378, 2011.

\bibitem{HASENBUSCH1994124}
Martin Hasenbusch, Mihai Marcu, and Klaus Pinn.
\newblock High precision renormalization group study of the roughening
  transition.
\newblock {\em Physica A: Statistical Mechanics and its Applications},
  208(1):124 -- 161, 1994.

\bibitem{Caselle:2006dv}
Michele Caselle, Martin Hasenbusch, and Marco Panero.
\newblock {High precision Monte Carlo simulations of interfaces in the
  three-dimensional ising model: A Comparison with the Nambu-Goto effective
  string model}.
\newblock {\em JHEP}, 03:084, 2006.

\bibitem{Bringoltz:2008nd}
Barak Bringoltz and Michael Teper.
\newblock {Closed k-strings in SU(N) gauge theories : 2+1 dimensions}.
\newblock {\em Phys. Lett.}, B663:429--437, 2008.

\bibitem{Athenodorou:2008cj}
Andreas Athenodorou, Barak Bringoltz, and Michael Teper.
\newblock {On the spectrum of closed k = 2 flux tubes in D=2+1 SU(N) gauge
  theories}.
\newblock {\em JHEP}, 05:019, 2009.

\bibitem{Juge:2002br}
K.~Jimmy Juge, Julius Kuti, and Colin Morningstar.
\newblock {Fine structure of the QCD string spectrum}.
\newblock {\em Phys. Rev. Lett.}, 90:161601, 2003.

\bibitem{HariDass:2006pq}
N.~D. Hari~Dass and Pushan Majumdar.
\newblock {String-like behaviour of 4-D SU(3) Yang-Mills flux tubes}.
\newblock {\em JHEP}, 10:020, 2006.

\bibitem{Giudice:2006hw}
Pietro Giudice, Ferdinando Gliozzi, and Stefano Lottini.
\newblock {Quantum broadening of k-strings in gauge theories}.
\newblock {\em JHEP}, 01:084, 2007.

\bibitem{Luscher:2004ib}
Martin Luscher and Peter Weisz.
\newblock {String excitation energies in SU(N) gauge theories beyond the
  free-string approximation}.
\newblock {\em JHEP}, 07:014, 2004.

\bibitem{Pepe:2010na}
M.~Pepe.
\newblock {String effects in Yang-Mills theory}.
\newblock {\em PoS}, LATTICE2010:017, 2010.

\bibitem{Luscher:1980iy}
M.~Luscher, G.~Munster, and P.~Weisz.
\newblock {How Thick Are Chromoelectric Flux Tubes?}
\newblock {\em Nucl. Phys.}, B180:1--12, 1981.

\bibitem{Caselle:2010zs}
M.~Caselle.
\newblock {Flux tube delocalization at the deconfinement point}.
\newblock {\em JHEP}, 08:063, 2010.

\bibitem{PhysRevD.82.094503}
A.~S. Bakry et~al.
\newblock String effects and the distribution of the glue in static mesons at
  finite temperature.
\newblock {\em Phys. Rev. D}, 82(9):094503, Nov 2010,hep-lat/1004.0782.

\bibitem{Bakry:2010sp}
Ahmed~S. Bakry, Derek~B. Leinweber, and Anthony~G. Williams.
\newblock {Bosonic stringlike behavior and the Ultraviolet filtering of QCD}.
\newblock {\em Phys.Rev.}, D85:034504, 2012.

\bibitem{Bakry:2011cn}
Ahmed~S. Bakry, Derek~B. Leinweber, and Anthony~G. Williams.
\newblock {On the ground state of Yang-Mills theory}.
\newblock {\em Annals Phys.}, 326:2165--2173, 2011.

\bibitem{Aharony:2010db}
Ofer Aharony and Nizan Klinghoffer.
\newblock Corrections to nambu-goto energy levels from the effective string
  action.
\newblock {\em Journal of High Energy Physics}, 2010(12):1--18, 2010.

\bibitem{Caselle:2014eka}
Michele Caselle, Marco Panero, Roberto Pellegrini, and Davide Vadacchino.
\newblock {A different kind of string}.
\newblock {\em JHEP}, 01:105, 2015.

\bibitem{Brandt:2017yzw}
Bastian~B. Brandt.
\newblock {Spectrum of the open QCD flux tube and its effective string
  description I: 3d static potential in SU(N = 2,3)}.
\newblock {\em JHEP}, 07:008, 2017.

\bibitem{Bakry:2017utr}
A.~Bakry, X.~Chen, M.~Deliyergiyev, A.~Galal, S.~Xu, and P.~M. Zhang.
\newblock {Stiff self-interacting string near QCD deconfinement point}.
\newblock 2017.

\bibitem{Khalil:2022idw}
M.~N. Khalil, A.~Bakry, X.~Chen, M.~Deliyergiyev, A.~Galal, A.~Khalaf, and
  P.~M. Zhang.
\newblock {On boundary corrections of L\"uscher-Weisz string}.
\newblock {\em EPJ Web Conf.}, 258:02004, 2022.

\bibitem{Bakry:2020ebo}
A.~S. Bakry, M.~A. Deliyergiyev, A.~A. Galal, A.~M. Khalaf, and M.~Khalil
  William.
\newblock {Quantum delocalization of strings with boundary action in Yang-Mills
  theory}.
\newblock 1 2020.

\bibitem{Bakry:2020flt}
A.~S. Bakry, M.~A. Deliyergiyev, A.~A. Galal, and M.~N. Khalil.
\newblock {On QCD strings beyond non-interacting model}.
\newblock 1 2020.

\bibitem{Billo:2011fd}
Marco Billo, Michele Caselle, and Roberto Pellegrini.
\newblock {New numerical results and novel effective string predictions for
  Wilson loops}.
\newblock {\em JHEP}, 01:104, 2012.
\newblock [Erratum: JHEP04,097(2013)].

\bibitem{nambu1974strings}
Yoichiro Nambu.
\newblock Strings, monopoles, and gauge fields.
\newblock {\em Physical Review D}, 10(12):4262, 1974.

\bibitem{Low:2001bw}
Ian Low and Aneesh~V. Manohar.
\newblock {Spontaneously broken space-time symmetries and Goldstone's theorem}.
\newblock {\em Phys. Rev. Lett.}, 88:101602, 2002.

\bibitem{Luscher:2002qv}
Martin Luscher and Peter Weisz.
\newblock {Quark confinement and the bosonic string}.
\newblock {\em JHEP}, 07:049, 2002.

\bibitem{Bakry:2017fii}
A.~S. Bakry, X.~Chen, M.~Deliyergiyev, A.~Galal, and P-M. Zhang.
\newblock {Perturbative width of open rigid-strings}.
\newblock 2017.

\bibitem{axion}
A~Bakry, M~Deliyergiyev, A~Galal, and M~Khalil Williams.
\newblock Smooth flux-sheets with topological winding modes.
\newblock {\em arXiv preprint arXiv:2005.04675}, 2020.

\bibitem{Arvis:1983fp}
J.~F. Arvis.
\newblock {The Exact $q \bar{q}$ Potential in Nambu String Theory}.
\newblock {\em Phys. Lett.}, 127B:106--108, 1983.

\bibitem{PhysRevD.27.2944}
K.~Dietz and T.~Filk.
\newblock Renormalization of string functionals.
\newblock {\em Phys. Rev. D}, 27(12):2944--2955, Jun 1983.

\bibitem{Caselle2010pf}
Michele Caselle and M~Zago.
\newblock A new approach to the study of effective string corrections in lgts.
\newblock {\em The European Physical Journal C}, 71(5):1--10, 2011.

\bibitem{Billo:2012da}
Marco Bill{\'o}, Michele Caselle, Ferdinando Gliozzi, M~Meineri, and Roberto
  Pellegrini.
\newblock The lorentz-invariant boundary action of the confining string and its
  universal contribution to the inter-quark potential.
\newblock {\em Journal of High Energy Physics}, 2012(5):1--18, 2012.

\bibitem{Bound}
AS~Bakry, MA~Deliyergiyev, AA~Galal, and M~Khalil Williams.
\newblock Boundary action and profile of effective bosonic strings.
\newblock {\em arXiv preprint arXiv:1912.13381}, 2019.

\bibitem{bakry2020qcd}
AS~Bakry, MA~Deliyergiyev, AA~Galal, and M~Khalil William.
\newblock On qcd strings beyond non-interacting model.
\newblock {\em arXiv preprint arXiv:2001.04203}, 2020.

\bibitem{allais}
A.~Allais and M.~Caselle.
\newblock {On the linear increase of the flux tube thickness near the
  deconfinement transition}.
\newblock {\em JHEP}, 01:073, 2009.

\bibitem{Gliozzi:2010zv}
F.~Gliozzi, M.~Pepe, and U.-J. Wiese.
\newblock {The Width of the Confining String in Yang-Mills Theory}.
\newblock {\em Phys.Rev.Lett.}, 104:232001, 2010.

\bibitem{Gliozzi:2010zt}
F.~Gliozzi, M.~Pepe, and U.~J. Wiese.
\newblock {The width of the color flux tube at 2-loop order}.
\newblock 2010.

\bibitem{Fabricius}
K.~Fabricius and O.~Haan.
\newblock {Heat bath method for the twisted Eguchi-Kawai model}.
\newblock {\em Phys. Lett}, B143:459, 1984.

\bibitem{Kennedy}
A.~D. Kennedy and B.~J. Pendleton.
\newblock {Improved heat bath method for Monte Carlo calculations in lattice
  gauge theories}.
\newblock {\em Phys. Lett}, B156:393--399, 1985.

\bibitem{Parisi}
G.~Parisi, R.~Petronzio, and F.~Rapuano.
\newblock {A measurement of the string tension near the continuum Limit}.
\newblock {\em Phys. Lett.}, B128:418, 1983.

\bibitem{1985PhLB15177D}
P.~{de Forcrand} and C.~{Roiesnel}.
\newblock {Refined methods for measuring large-distance correlations}.
\newblock {\em Phys. Lett. B}, 151:77--80, January 1985.

\bibitem{Albanese}
M.~Albanese et~al.
\newblock {Glueball masses and string tension in lattice QCD}.
\newblock {\em Phys. Lett}, B192:163--169, 1987.

\bibitem{Gattringer}
Christof Gattringer and Christian Lang.
\newblock {\em Quantum chromodynamics on the lattice: an introductory
  presentation}, volume 788.
\newblock Springer Science \& Business Media, 2009.

\end{thebibliography}


\begin{thebibliography}{99}

\bibitem{Creutz:1980zw}
M.~Creutz.
\newblock {Monte carlo study of quantized SU(2) gauge theory}.
\newblock {\em Phys. Rev.}, \textbf{D21}:2308--2315, 1980.

\bibitem{Bali:1994de}
G.~S. Bali, K.~Schilling, and C.~Schlichter.
\newblock {Observing long color flux tubes in SU(2) lattice gauge theory}.
\newblock {\em Phys. Rev.}, D51:5165--5198, 1995.

\bibitem{Haymaker:1994fm}
Richard~W. Haymaker, Vandana Singh, Yingcai Peng, and Jacek Wosiek.
\newblock Distribution of the color fields around static quarks.
\newblock {\em Physical Review D}, 53(1):389, 1996.

\bibitem{Cea:1995zt}
Paolo Cea and Leonardo Cosmai.
\newblock Dual superconductivity in the SU(2) pure gauge vacuum: A lattice
  study.
\newblock {\em Physical Review D}, 52(9):5152, 1995.

\bibitem{Okiharu:2003vt}
F.~Okiharu and R. M.~Woloshyn.
\newblock A study of colour field distributions in the baryon.
\newblock {\em Nuclear Physics B-Proceedings Supplements}, 129:745--747, 2004.

\bibitem{Bakry:2017jna}
A.~Bakry, X.~Chen, M.~Deliyergiyev, A.~Galal, S.~Xu, and P.~M. Zhang.
\newblock {Mesonic String of Diquark-Quark Configuration at Finite
  Temperature}.
\newblock {\em PoS}, Hadron2017:211, 2018.

\bibitem{Bakry:2014gea}
Ahmed~S. Bakry, Xurong Chen, and Peng-Ming Zhang.
\newblock {Y-stringlike behavior of a static baryon at finite temperature}.
\newblock {\em Phys. Rev.}, D91:114506, 2015.

\bibitem{Bakry:2020xcu}
A.~S. Bakry, M.~A. Deliyergiyev, A.~A. Galal, and A.~M. Khalaf.
\newblock {Strings of diquark-quark (QQ)Q baryon before phase transition}.
\newblock {\em arXiv/heplat:2002.07202}.

\bibitem{PhysRevB.78.024510}
Mark~G. Alford and Gerald Good.
\newblock Flux tubes and the type-i/type-ii transition in a superconductor
  coupled to a superfluid.
\newblock {\em Phys. Rev. B}, 78(2):024510, Jul 2008.

\bibitem{2007arXiv0709.1042K}
K.~{Kasamatsu} and M.~{Tsubota}.
\newblock {Quantized vortices in atomic Bose-Einstein condensates}.
\newblock {\em ArXiv e-prints}, September 2007.

\bibitem{Nielsen197345}
H. B. Nielsen and P.~Olesen.
\newblock Vortex-line models for dual strings.
\newblock {\em Nuclear Physics B}, 61:45--61, 1973.

\bibitem{Lo:2005xt}
Amy~S. Lo and Edward~L. Wright.
\newblock {Signatures of cosmic strings in the cosmic microwave background}.
\newblock 2005.

\bibitem{PhysRev.177.2247}
Curtis~G. Callan~Jr., Sidney Coleman, Julius Wess, and Bruno Zumino.
\newblock Structure of phenomenological lagrangians. ii.
\newblock {\em Physical Review}, 177(5):2247, 1969.

\bibitem{Luscher:2002qv}
Martin Luscher and Peter Weisz.
\newblock {Quark confinement and the bosonic string}.
\newblock {\em JHEP}, 07:049, 2002.

\bibitem{Juge:2002br}
K.~Jimmy Juge, Julius Kuti, and Colin Morningstar.
\newblock {Fine structure of the QCD string spectrum}.
\newblock {\em Phys. Rev. Lett.}, 90:161601, 2003.

\bibitem{HariDass2008273}
N. D.~Hari Dass and Pushan Majumdar.
\newblock Continuum limit of string formation in 3d SU(2) lgt.
\newblock {\em Phys. Lett. B}, 658(5):273--278, 2008.

\bibitem{Caselle:2016mqu}
Michele Caselle, Marco Panero, and Davide Vadacchino.
\newblock {Width of the flux tube in compact U(1) gauge theory in three
  dimensions}.
\newblock {\em JHEP}, 02:180, 2016.

\bibitem{caselle-2002}
M.~Caselle, M.~Panero, P.~Provero, and M.~Hasenbusch.
\newblock {String effects in Polyakov loop correlators}.
\newblock {\em Nucl. Phys. Proc. Suppl.}, 119:499--501, 2003.

\bibitem{Pennanen:1997qm}
P.~Pennanen, Anthony~M. Green, and Christopher Michael.
\newblock {Flux-tube structure and beta-functions in SU(2)}.
\newblock {\em Phys. Rev.}, D56:3903--3916, 1997.

\bibitem{Brandt:2016xsp}
Bastian~B. Brandt and Marco Meineri.
\newblock {Effective string description of confining flux tubes}.
\newblock {\em Int. J. Mod. Phys.}, A31(22):1643001, 2016.

\bibitem{Luscher:1980iy}
M.~Luscher, G.~Munster, and P.~Weisz.
\newblock {How Thick Are Chromoelectric Flux Tubes?}
\newblock {\em Nucl. Phys.}, B180:1--12, 1981.

\bibitem{Caselle:1995fh}
M.~Caselle, F.~Gliozzi, Ulrika Magnea, and S.~Vinti.
\newblock {Width of long colour flux tubes in lattice gauge systems}.
\newblock {\em Nucl. Phys.}, B460:397--412, 1996.

\bibitem{Bonati:2011nt}
Claudio Bonati.
\newblock {Finite temperature effective string corrections in (3+1)D SU(2)
  lattice gauge theory}.
\newblock {\em Phys.Lett.}, B703:376--378, 2011.

\bibitem{HASENBUSCH1994124}
Martin Hasenbusch, Mihai Marcu, and Klaus Pinn.
\newblock High precision renormalization group study of the roughening
  transition.
\newblock {\em Physica A: Statistical Mechanics and its Applications},
  208(1):124--161, 1994.

\bibitem{Caselle:2006dv}
Michele Caselle, Martin Hasenbusch, and Marco Panero.
\newblock {High precision Monte Carlo simulations of interfaces in the
  three-dimensional ising model: A Comparison with the Nambu-Goto effective
  string model}.
\newblock {\em JHEP}, 03:084, 2006.

\bibitem{Bringoltz:2008nd}
Barak Bringoltz and Michael Teper.
\newblock {Closed k-strings in SU(N) gauge theories: 2+1 dimensions}.
\newblock {\em Phys. Lett.}, B663:429--437, 2008.

\bibitem{Athenodorou:2008cj}
Andreas Athenodorou, Barak Bringoltz, and Michael Teper.
\newblock {On the spectrum of closed k = 2 flux tubes in D=2+1 SU(N) gauge
  theories}.
\newblock {\em JHEP}, 05:019, 2009.

\bibitem{HariDass:2006pq}
N.~D. Hari~Dass and Pushan Majumdar.
\newblock {String-like behaviour of 4-D SU(3) Yang-Mills flux tubes}.
\newblock {\em JHEP}, 10:020, 2006.

\bibitem{Giudice:2006hw}
Pietro Giudice, Ferdinando Gliozzi, and Stefano Lottini.
\newblock {Quantum broadening of k-strings in gauge theories}.
\newblock {\em JHEP}, 01:084, 2007.

\bibitem{Luscher:2004ib}
Martin Luscher and Peter Weisz.
\newblock {String excitation energies in SU(N) gauge theories beyond the
  free-string approximation}.
\newblock {\em JHEP}, 07:014, 2004.

\bibitem{Caselle:2010zs}
M.~Caselle.
\newblock {Flux tube delocalization at the deconfinement point}.
\newblock {\em JHEP}, 08:063, 2010.

\bibitem{PhysRevD.82.094503}
A.~S. Bakry et~al.
\newblock String effects and the distribution of the glue in static mesons at
  finite temperature.
\newblock {\em Phys. Rev. D}, 82(9):094503, Nov 2010, hep-lat/1004.0782.

\bibitem{Bakry:2010sp}
Ahmed~S. Bakry, Derek~B. Leinweber, and Anthony~G. Williams.
\newblock {Bosonic stringlike behavior and the Ultraviolet filtering of QCD}.
\newblock {\em Phys.Rev.}, D85:034504, 2012.

\bibitem{Caselle:2004jq}
M.~Caselle, M.~Hasenbusch, and M.~Panero.
\newblock {Short distance behavior of the effective string}.
\newblock {\em JHEP}, 05:032, 2004.

\bibitem{Caselle:2004er}
M.~Caselle, M.~Pepe, and A.~Rago.
\newblock {Static quark potential and effective string corrections in the
  (2+1)-d SU(2) Yang-Mills theory}.
\newblock {\em JHEP}, 10:005, 2004.

\bibitem{Ambjorn:2014rwa}
J.~Ambjørn, Y.~Makeenko, and A.~Sedrakyan.
\newblock {Effective QCD string beyond the Nambu-Goto action}.
\newblock {\em Phys. Rev.}, D89(10):106010, 2014.

\bibitem{Caselle:2014eka}
Michele Caselle, Marco Panero, Roberto Pellegrini, and Davide Vadacchino.
\newblock {A different kind of string}.
\newblock {\em JHEP}, 01:105, 2015.

\bibitem{bakry2020qcd}
A. S.~Bakry, M. A.~Deliyergiyev, A. A.~Galal, and M.~Khalil William.
\newblock On qcd strings beyond non-interacting model.
\newblock {\em arXiv preprint arXiv:2001.04203 [hep-lat]}, 2020.

\bibitem{Bakry:2020ebo}
A.~S. Bakry, M.~A. Deliyergiyev, A.~A. Galal, A.~M. Khalaf, and M.~Khalil
  William.
\newblock {Quantum delocalization of strings with boundary action in Yang-Mills
  theory}.
\newblock {\em arXiv preprint arXiv:2001.02392 [hep-lat]}, 2020.
\bibitem{Brandt:2017yzw}
Bastian~B. Brandt.
\newblock {Spectrum of the open QCD flux tube}.
\newblock {\em JHEP}, 07:008, 2017.

\bibitem{Aharony:2010db}
Ofer Aharony and Nizan Klinghoffer.
\newblock Corrections to nambu-goto energy levels from the effective string
  action.
\newblock {\em Journal of High Energy Physics}, 2010(12):1--18, 2010.

\bibitem{Bakry:2020flt}
A.~S. Bakry, M.~A. Deliyergiyev, A.~A. Galal, and M.~N. Khalil.
\newblock {On QCD strings beyond non-interacting model}.

\bibitem{Billo:2011fd}
Marco Billo, Michele Caselle, and Roberto Pellegrini.
\newblock {New numerical results and novel effective string predictions for
  Wilson loops}.
\newblock {\em JHEP}, 01:104, 2012.
\newblock [Erratum: JHEP04,097(2013)].

\bibitem{Brandt:2010bw}
Bastian~B. Brandt.
\newblock {Probing boundary-corrections to Nambu-Goto open string energy levels
  in 3d SU(2) gauge theory}.
\newblock {\em JHEP}, 02:040, 2011.

\bibitem{Bound}
A. S.~Bakry, M. A.~Deliyergiyev, A. A.~Galal, and M.~Khalil Williams.
\newblock Boundary action and profile of effective bosonic strings.
\newblock {\em arXiv preprint arXiv:1912.13381}, 2019.

\bibitem{Billo:2012da}
Marco Bill{\'o}, Michele Caselle, Ferdinando Gliozzi, M.~Meineri, and Roberto
  Pellegrini.
\newblock The lorentz-invariant boundary action of the confining string and its
  universal contribution to the inter-quark potential.
\newblock {\em Journal of High Energy Physics}, 2012(5):1--18, 2012.

\bibitem{nambu1974strings}
Yoichiro Nambu.
\newblock Strings, monopoles, and gauge fields.
\newblock {\em Physical Review D}, 10(12):4262, 1974.

\bibitem{Low:2001bw}
Ian Low and Aneesh~V. Manohar.
\newblock {Spontaneously broken space-time symmetries and Goldstone's theorem}.
\newblock {\em Phys. Rev. Lett.}, 88:101602, 2002.

\bibitem{Aharony:2009gg}
Ofer Aharony and Eyal Karzbrun.
\newblock {On the effective action of confining strings}.
\newblock {\em JHEP}, 06:012, 2009.

\bibitem{Aharony2011}
Ofer Aharony and Matan Field.
\newblock On the effective theory of long open strings.
\newblock {\em Journal of High Energy Physics}, 2011(1):65, Jan 2011.

\bibitem{Dubovsky:2012wk}
Sergei Dubovsky, Raphael Flauger, and Victor Gorbenko.
\newblock {Solving the Simplest Theory of Quantum Gravity}.
\newblock {\em JHEP}, 09:133, 2012.

\bibitem{Billo:2010ix}
Marco Billo, Michele Caselle, Valentina Verduci, and Mirco Zago.
\newblock {New results on the effective string corrections to the inter-quark
  potential}.
\newblock {\em PoS}, LATTICE2010:273, 2010.

\bibitem{Bakry:2017fii}
A.~S. Bakry, M.~A. Deliyergiyev, A.~A. Galal, and M.~N. Khalil.
\newblock {Perturbative width of open rigid strings}.
\newblock {\em Phys. Rev. D}, 105(9):094504, 2022.

\bibitem{axion}
A.~Bakry, M.~Deliyergiyev, A.~Galal, and M.~Khalil Williams.
\newblock Smooth flux-sheets with topological winding modes.
\newblock {\em arXiv preprint arXiv:2005.04675}, 2020.

\bibitem{Arvis:1983fp}
J.~F. Arvis.
\newblock {The Exact $q \bar{q}$ Potential in Nambu String Theory}.
\newblock {\em Phys. Lett.}, 127B:106--108, 1983.

\bibitem{POLYAKOV19}
A.~Polyakov.
\newblock Fine structure of strings.
\newblock {\em Nuclear Physics B}, 268(2):406--412, 1986.

\bibitem{Kleinert:1986bk} H.~Kleinert,
{\textit{The Membrane Properties of Condensing Strings}},
Phys. Lett. {\bf{B174}} (1986) 335--338.

\bibitem{German:1989vk}
G.~German and H.~Kleinert.
\newblock {Perturbative Two Loop Quark Potential of Stiff Strings in Any
  Dimension}.
\newblock {\em Phys. Rev.}, D40:1108--1119, 1989.

\bibitem{Viswanathan:1988ad}
K.~S. Viswanathan and Xiao-An Zhou.
\newblock {FREE ENERGY AND THE STATIC POTENTIAL FOR OPEN SMOOTH STRINGS}.
\newblock {\em Int. J. Mod. Phys.}, A3:2195--2206, 1988.

\bibitem{PhysRevD.27.2944}
K.~Dietz and T.~Filk.
\newblock Renormalization of string functionals.
\newblock {\em Phys. Rev. D}, 27(12):2944--2955, Jun 1983.

\bibitem{Hellerman:2014cba}
Simeon Hellerman, Shunsuke Maeda, Jonathan Maltz, and Ian Swanson.
\newblock {Effective String Theory Simplified}.
\newblock {\em JHEP}, 09:183, 2014.

\bibitem{Giudice:2009di}
Pietro Giudice, Ferdinando Gliozzi, and Stefano Lottini.
\newblock {The Confining string beyond the free-string approximation in the
  gauge dual of percolation}.
\newblock {\em JHEP}, 03:104, 2009.

\bibitem{Aharony:2013ipa}
Ofer Aharony and Zohar Komargodski.
\newblock {The Effective Theory of Long Strings}.
\newblock {\em JHEP}, 05:118, 2013.

\bibitem{Dubovsky:2012sh}
Sergei Dubovsky, Raphael Flauger, and Victor Gorbenko.
\newblock {Effective String Theory Revisited}.
\newblock {\em JHEP}, 09:044, 2012.

\bibitem{Dubovsky:2014fma}
Sergei Dubovsky, Raphael Flauger, and Victor Gorbenko.
\newblock {Flux Tube Spectra from Approximate Integrability at Low Energies}.
\newblock {\em J. Exp. Theor. Phys.}, 120:399--422, 2015.

\bibitem{Caselle2010pf}
Michele Caselle and M.~Zago.
\newblock A new approach to the study of effective string corrections in lgts.
\newblock {\em The European Physical Journal C}, 71(5):1--10, 2011.

\bibitem{abramowitz}
I. A. Stegun and M. Abramowitz.
\newblock {\em Handbook of Mathematical Functions with Formulas, Graphs, and
  Mathematical Tables}, volume~1.
\newblock (Dover, New York), ninth dover printing, tenth gpo printing edition.,
  1964.

\bibitem{allais}
A.~Allais and M.~Caselle.
\newblock {On the linear increase of the flux tube thickness near the
  deconfinement transition}.
\newblock {\em JHEP}, 01:073, 2009.

\bibitem{Gliozzi:2010zv}
F.~Gliozzi, M.~Pepe, and U.-J. Wiese.
\newblock {The Width of the Confining String in Yang-Mills Theory}.
\newblock {\em Phys.Rev.Lett.}, 104:232001, 2010.

\bibitem{Gliozzi:2010zt}
F.~Gliozzi, M.~Pepe, and U.~J. Wiese.
\newblock {The width of the color flux tube at 2-loop order}.
\newblock {\em JHEP}, 11:053, 2010.

\bibitem{Gattringer}
C.~Gattringer and C.~Lang.
\newblock {\em Quantum chromodynamics on the lattice}, Vol 788.
\newblock Springer Science \& Business Media, 2009.

\end{thebibliography}

\end{document}